%% file: tit-mimo-asympt-fairness-v1.3.tex
\documentclass[11pt,draftclsnofoot,peerreviewca,letterpaper,onecolumn]{IEEEtran}
\usepackage{cite}
\usepackage{graphicx,color,epsfig,rotating}
\usepackage{amsfonts,amsmath,amssymb}
\usepackage{algorithm,algorithmic}
\usepackage{subfigure}
\input{macros}

\newcommand{\MMSE}{{\sf mmse}}

\newtheorem{theorem}{Theorem}
\newtheorem{lemma}{Lemma}

\begin{document}

\title{Multi-Cell MIMO Downlink \\
    with Cell Cooperation and Fair Scheduling: \\
    a Large-System Limit Analysis}

\author{Hoon Huh,~\IEEEmembership{Student Member,~IEEE},
Giuseppe Caire,~\IEEEmembership{Fellow,~IEEE}, \\
Sung-Hyun Moon,~\IEEEmembership{Student Member,~IEEE},
Young-Tae Kim,~\IEEEmembership{Student Member,~IEEE}, \\
and Inkyu Lee,~\IEEEmembership{Senior Member,~IEEE}
\thanks{H. Huh and G. Caire are with the Department of Electrical Engineering,
University of Southern California, Los Angeles, CA, USA. (e-mail: hhuh, caire@usc.edu)}
\thanks{S.-H. Moon, Y.-T. Kim and I. Lee are with the School of Electrical Engineering,
Korea University, Seoul, Korea. (e-mail: shmoon, reftm, inkyu@korea.ac.kr)}
\thanks{The material in this paper was presented in part at the 2010 IEEE International
Communications Conference (ICC), Cape Town, South Africa, May 2010 and will be
presented in part at the 2010 International Symposium on Information Theory (ISIT),
Austin, Texas, June 2010.}
}

\thispagestyle{empty}

\maketitle

\begin{abstract}
We consider the downlink of a cellular network with multiple cells and multi-antenna base
stations, including a realistic distance-dependent pathloss model, clusters of cooperating
cells, and general ``fairness'' requirements. Beyond Monte Carlo simulation, no efficient
computation method to evaluate the ergodic throughput of such systems has been presented
so far. We propose an analytic solution based on the combination of large random matrix results
and convex optimization. The proposed method is computationally much more efficient than
Monte Carlo simulation and provides surprisingly accurate approximations for the actual
finite-dimensional systems, even for a small number of users and base station antennas.
Numerical examples include 2-cell linear and three-sectored 7-cell planar layouts,
with no inter-cell cooperation, sector cooperation, or full inter-cell cooperation.
\end{abstract}

\begin{IEEEkeywords}
Asymptotic analysis, fairness scheduling, inter-cell cooperation, large-system limit,
multi-cell MIMO downlink, weighted sum rate maximization.
\end{IEEEkeywords}

\newpage
\setcounter{page}{1}

\section{Introduction} \label{sec:intro}

\IEEEPARstart{M}{ultiuser} MIMO (MU-MIMO) technology is expected to play a key role in future
wireless cellular networks \cite{IEEE80216m, LTE-Advanced}. The MIMO Gaussian
Broadcast Channel (BC) model
\cite{Caire-Shamai-TIT03,Viswanath-Tse-TIT03,Vishwanath-Jindal-Goldsmith-TIT03,
Yu-Cioffi-TIT04,Weingarten-Steinberg-Shamai-TIT06} serves as the information theoretic foundation for
various MU-MIMO downlink schemes. In particular, the MIMO Gaussian BC  capacity region
with zero common message rate was characterized in \cite{Weingarten-Steinberg-Shamai-TIT06} where
the optimality of Gaussian Dirty-Paper Coding (DPC) is shown, subject to a general
convex input covariance  constraint.

In a multi-cell scenario, depending on the level of inter-cell cooperation, we are in the presence
of a MIMO broadcast and interference channel, which is not yet fully understood in an information
theoretic sense. A simple and analytically tractable model for the multi-cell system was introduced
by Wyner \cite{Wyner-TIT94}. In both one-dimensional linear cellular array and two-dimensional
hexagonal cellular pattern, only interference from adjacent cells is considered with a single
scaling factor, and the uplink capacity is obtained in a closed form in the case of full joint processing of
all cells and no fading. Wyner's setting was extended in several works \cite{Shamai-Wyner-TIT97,
Somekh-Shamai-TIT00, Somekh-Shamai-TIT07, Shamai-PIMRC08, Sanderovich-TIT09}. Single cell processing
and joint two-cell processing was investigated by treating the inter-cell interference as Gaussian
noise in \cite{Shamai-Wyner-TIT97} and the flat-fading channel case with full joint cell processing
was treated in \cite{Somekh-Shamai-TIT00}. This model was modified and extended to take into account
various issues such as soft hand-off and limited inter-cell cooperation due to constrained backhaul
capacity (see \cite{Somekh-Shamai-TIT07, Shamai-PIMRC08, Sanderovich-TIT09} and references therein).

Although the Wyner model captures some fundamental aspects of the multi-cell problem, its rather
unrealistic assumption for the pathloss makes the system essentially symmetric with respect to any user.
More realistically, users in different locations of the cellular coverage region are subject to
distance-dependent pathloss that may have more than 30 dB of dynamic range \cite{Wimax-eval06},
and therefore they are in fundamentally asymmetric conditions.
It follows that characterizing the sum-capacity (or
achievable sum-throughput, under some suboptimal scheme)
is rather meaningless from a system performance viewpoint, unless some appropriate notion of
{\em fairness} is taken into account. In fact, if the sum-throughput is the only objective, the resulting
rate and power optimization under distance-dependent pathloss
would lead to the solution of serving only the users close to their base station (BS), while leaving the users at the cell edge
to starve.

As a matter of fact,  ``fairness'' is a fundamental aspect in cellular networks.
The problem of {\em downlink scheduling} subject to some fairness criterion
has been widely studied (see for example \cite{Viswanath-Tse-Laroia-TIT02,
Georgiadis-Neely-Tassiulas-04, ShiraniMehr-Caire-Neely-submit09} and references therein).
The goal of fairness scheduling is to make the system operate at some point of its ergodic achievable
rate region such that a suitable concave and increasing {\em network utility function} is maximized
\cite{Mo-TNET00}. By choosing the shape of the network utility function, a desired fairness
criterion can be enforced. The framework of {\em stochastic network optimization}
\cite{Georgiadis-Neely-Tassiulas-04} can be leveraged in order to systematically devise scheduling
algorithms that perform arbitrarily close to the optimal achievable fairness point, even when the
explicit computation of the achievable ergodic rate region is hopelessly complicated.
The fairness operating point is given as the time-averaged rate obtained by applying a dynamic scheduling
algorithm on a slot-by-slot basis. Hence, its analytical characterization is generally very difficult and the
system performance is typically evaluated by letting the scheduling algorithm evolve in time and
computing the time-averaged rates by Monte Carlo simulation \cite{Huang-PIMRC05, Boccardi-Huang-PIMRC07,
Zhang-Dai-TWC08, Caire-Docomo-Allerton08, Marsch-GC08, Zhang-Heath-TWC09, Huang-Valenzuela-TWC09,
Ramprashad-Caire-PIMRC09, Parkvall-Zangi-etal-VTC08F, Landre-PIMRC09, Farajidana-GC09}.

In this paper, we propose an alternative approach based on the ``large-system limit.'' We leverage
results on large random matrices \cite{Tulino-04, Girko-90, Tulino-TIT05, Tulino-TWC06, Aktas-TIT06},
in order to characterize the system achievable rate region in the limit where both the number of
antennas per BS and the number of users per cell grow to infinity with a fixed ratio. Our model
encompasses arbitrary user locations and distance-dependent pathloss and considers arbitrary inter-cell
cooperation clusters, where the BSs in the same cluster operate as a distributed antenna array (full
cooperation) and inter-cluster interference is treated as Gaussian noise (no inter-cluster cooperation).
As special cases, we recover conventional cellular systems (no inter-cell cooperation) and the case
of full cooperation. In the large-system limit, the channel randomness disappears and the MU-MIMO system
becomes a deterministic network. It follows that the performance of dynamic fairness scheduling can be
calculated by solving a ``static'' convex optimization problem. By incorporating the large random matrix
results into the convex optimization solution, we solve this problem in almost closed form
(up to the numerical solution of a fixed-point equation). The solution is particularly simple when each cooperation
cluster satisfies certain symmetry conditions that will be discussed later on.
The proposed method is much more efficient than Monte Carlo simulation and, somehow surprisingly,
it provides results that match very closely the performance
of finite-dimensional systems, even for very small dimension.

The remainder of this paper is organized as follows. In Section \ref{sec:setup}, we present the MU-MIMO
downlink system model with cell cluster cooperation and formulate the fairness scheduling problem. We
develop the numerical solution for the input covariance maximizing the weighted average sum rate in the
large-system limit in Section \ref{sec:wesrm}. In Section \ref{sec:fairness}, we use these results
in order to obtain a semi-analytic method to calculate the optimal ergodic fairness rate point in the
asymptotic regime. In Section \ref{sec:result}, the asymptotic rates are shown in 2-cell linear and
7-cell planar models and are compared with finite-dimensional simulation results obtained by the
combination of DPC and the actual dynamic scheduling scheme based on stochastic optimization.
Concluding remarks are presented in Section \ref{sec:conclusion}.

\section{Problem setup} \label{sec:setup}

We consider $M$ BSs with $\gamma N$ antennas each, and $KN$ single-antenna user terminals, distributed
in the cellular coverage region. Users are divided into $K$ co-located ``user groups'' of equal size $N$.
Users in the same group are statistically equivalent: they experience the same pathloss from all BSs
and their small-scale fading channel coefficients are independent and identically-distributed (i.i.d.).
In practice, it is reasonable to assume that co-located users are separated by a sufficient number of
wavelength such that they undergo i.i.d. small-scale fading, but the wavelength is sufficiently small
so that they all have essentially the same distance-dependent pathloss. Users in different groups
observe generally different pathlosses, depending on their relative positions with respect to the BSs.

We assume a block-fading model where the channel coefficients are constant over time-frequency ``slots''
determined by the channel coherence time and bandwidth, and change according to some well-defined ergodic
process from slot to slot. In contrast, the distance-dependent pathloss coefficients are constant in time.
This is representative of a typical situation where the distance between BSs and users changes
significantly over a time-scale of the order of tens of seconds, while the small-scale fading
decorrelates completely within a few milliseconds \cite{Tse-05}. The slot index shall be denoted by $t$,
but we will omit $t$ for notation simplicity whenever possible. We shall make explicit reference to the
time slot when discussing the dynamic fairness scheduling policy in Section \ref{sec:fairness}.

One channel use of the multi-cell MU-MIMO downlink is described by
\begin{equation} \label{eq:ch_model}
\yv_k = \sum_{m=1}^{M} \alpha_{m,k} \Hm_{m,k}^\herm \xv_m + \nv_k
\end{equation}
where $\yv_k = [y_{k,1}\ldots y_{k,N}]^\transp \in \CC^N$ denotes the received signal vector for the
$k$-th user group, $\alpha_{m,k}$ and $\Hm_{m,k}$ denote the the distance-dependent pathloss and a
$\gamma N \times N$ channel matrix collecting the small-scale channel fading coefficients from the
$m$-th BS to the $k$-th user group, respectively,
$\xv_m = [x_{m,1}\ldots x_{m,\gamma N}]^\transp \in\CC^{\gamma N}$ is the signal vector transmitted by
the $m$-th BS, and $\nv_k=[n_{k,1}\ldots n_{k,N}]^\transp \in \CC^N$ denotes the AWGN at the user
receivers in the $k$-th user group. The elements of $\nv_k$ and of $\Hm_{m,k}$ are i.i.d.
$\sim \Cc\Nc(0,1)$. We assume a per-BS average power constraint expressed by
$\trace \left( {\rm Cov}(\xv_m) \right) \leq P_m$, where $P_m > 0$ denotes the total transmit power of
the $m$-th BS.

We assume that the BSs are grouped into cooperation clusters. Each cluster acts effectively as a
distributed MU-MIMO system, with a distributed transmit antenna array formed by all antennas of all BS
in the cluster. Each cluster has perfect channel state information for all the users associated
with the cluster, and has {\em statistical information} (i.e., known distributions but not the
instantaneous values) relative to signals from other clusters. Within these channel state information
assumptions, we consider ideal joint processing of all BSs in the same cluster. Inter-Cluster
Interference (ICI) is treated as additional Gaussian noise.
Let $L$ denote the number of cooperation clusters. We define the BS partition
$\{\Mc_1,\ldots,\Mc_L\}$ of the set $\{1,\ldots, M\}$ and the corresponding user group partition
$\{\Kc_1,\ldots, \Kc_L\}$ of the set $\{1,\ldots, K\}$, where $\Mc_\ell$ and $\Kc_\ell$ denote the set
of BSs and user groups forming the $\ell$-th cooperation cluster. We assume that clusters are selfish
and use all available transmit power, with no consideration for the ICI that they may cause to other
clusters. Hence, the ICI plus noise variance at any user terminal in group $k \in \Kc_\ell$ is given by
\begin{align} \label{eq:ici-pow}
\sigma_k^2 & \;=\; \EE \left[ \frac{1}{N} \left\| \sum_{m \notin \Mc_\ell} \alpha_{m,k}
  \Hm_{m,k}^\herm \xv_m + \nv_k \right\|^2 \right] \nonumber \\
& \;=\; 1 + \sum_{m \notin \Mc_\ell} \alpha_{m,k}^2 P_m.
\end{align}
From the viewpoint of cluster $\Mc_\ell$, the system is equivalent to a single-cell MU-MIMO downlink
with per-group-of-antennas power constraint where each antenna group corresponds to each BS's antennas,
and with AWGN power at the user receivers given by (\ref{eq:ici-pow}).
Therefore, from now on, we shall focus on a reference cluster (say, $\ell$) and simplify our notation.
We let $B = |\Mc_\ell|$ and $A = |\Kc_\ell|$ denote the number of BS and user groups in the cluster,
and enumerate the BS and the user groups forming the cluster as $m = 1,\ldots, B$ and $k = 1,\ldots, A$,
respectively. Also, we define the modified path coefficients
$\beta_{m,k} = \frac{\alpha_{m,k}}{\sigma_k}$ and the cluster channel matrix
\begin{equation} \label{eq:H}
\widetilde{\Hm} = \left[\begin{array}{ccc}
    \beta_{1,1} \Hm_{1,1} & \cdots & \beta_{1,A} \Hm_{1,A} \\
    \vdots                & \ddots & \vdots \\
    \beta_{B,1} \Hm_{B,1} & \cdots & \beta_{B,A} \Hm_{B,A}
    \end{array}\right].
\end{equation}
Hence, one channel use of the reference cluster downlink is given by
\begin{equation} \label{eq:single-bc}
\yv  = \widetilde{\Hm}^\herm \xv + \vv
\end{equation}
where $\yv = \CC^{A N}$, $\xv = \CC^{\gamma B N}$, and $\vv \sim \Cc\Nc(\zerov, \Id)$ (we drop subscript
$\ell$ for notation simplicity).

It is well-known that the boundary of the capacity region of the MIMO BC (\ref{eq:single-bc}) for fixed
channel matrix $\widetilde{\Hm}$ and given per-group-of-antennas power constraints $\{P_1,\ldots, P_B\}$
can be characterized by the solution of a min-max weighted sum-rate problem \cite{Yu-TIT06,
Zhang-Poor-etal-ISIT09, Huh-Caire-submitTSP09}. For reasons that will be clear when discussing the
scheduling policy in Section \ref{sec:fairness}, we restrict ourselves to the case of identical weights
for all statistically equivalent users, i.e., for the case that users in the same group have the same
weight for their individual rates. We let $W_k$ and
$R_k(\widetilde{\Hm}) = \frac{1}{N} \sum_{i=1}^N R_{k,i} (\widetilde{\Hm})$ denote the weight for user
group $k$ and the corresponding instantaneous per-user rate, respectively. In this paper, we refer to as
``instantaneous'' the quantities that depend on the realization of the channel matrix $\widetilde{\Hm}$.
Since this changes from slot to slot, instantaneous quantities also change accordingly.
We let $\pi$ denote the permutation that sorts the weights in increasing order
$W_{\pi_1} \leq \ldots \leq W_{\pi_A}$ and use the subscript $[k:A]$ to indicate quantities involving
user groups from $\pi_k$ to $\pi_A$. In particular, we let
$\widetilde{\Hm}_{k:A} = \left[ \widetilde{\Hm}_{\pi_k} \ldots \widetilde{\Hm}_{\pi_A} \right]$
and $\Qm_{k:A} = \diag \left( \Qm_{\pi_k}, \ldots, \Qm_{\pi_A} \right)$,
where $\widetilde{\Hm}_{k}$ is the $k$-th $\gamma B N \times N$ slice of $\widetilde{\Hm}$ in
(\ref{eq:H}), and where $\Qm_k = \diag(q_{k,1}, \ldots, q_{k,N})$ is a $N \times N$ non-negative
definite diagonal matrix.

The rate point $\{R_1(\widetilde{\Hm}), \ldots, R_A(\widetilde{\Hm})\}$ corresponding to weights
$\{W_1, \ldots, W_A\}$ is obtained as solution of the max-min problem \cite{Yu-TIT06,
Zhang-Poor-etal-ISIT09, Huh-Caire-submitTSP09}
\begin{equation} \label{eq:min-max}
\min_{\lambdav \geq 0} \; \max_{\Qm \geq 0} \;\; \sum_{k=1}^A W_{\pi_k} R_{\pi_k}(\widetilde{\Hm})
\end{equation}
for the instantaneous per-user rate of each group
\begin{equation} \label{eq:min-max-1}
R_{\pi_k}(\widetilde{\Hm}) =  \frac{1}{N} \log \frac{\left | \Sigmam(\lambdav) +
  \widetilde{\Hm}_{k:A} \Qm_{k:A} \widetilde{\Hm}_{k:A}^\herm \right |}
  {\left | \Sigmam(\lambdav) + \widetilde{\Hm}_{k+1:A} \Qm_{k+1:A}
  \widetilde{\Hm}_{k+1:A}^\herm \right |}
\end{equation}
where $\Sigmam(\lambdav)$ is a $\gamma B N \times \gamma B N$ block-diagonal matrix with
$\gamma N \times \gamma N$ constant diagonal blocks $\lambda_m \Id_{\gamma N}$, for $m = 1,\ldots, B$
and the maximization with respect to $\Qm$ is subject to the trace constraint
\begin{equation} \label{eq:trace-constr}
\trace (\Qm) \leq \sum_{m=1}^B \lambda_m P_m.
\end{equation}
The variables $\lambdav = \{\lambda_m\}$ are the Lagrange multipliers corresponding to the
per-group-of-antennas power constraints. The rate $R_{\pi_k}(\widetilde{\Hm})$ in (\ref{eq:min-max-1})
can be interpreted as the instantaneous per-user rate of user group $\pi_k$ in the dual vector Multiple
Access Channel (MAC) with worst-case noise defined by
\begin{equation} \label{eq:dual-mac}
\rv = \sum_{k=1}^A \widetilde{\Hm}_k \sv_k + \zv
\end{equation}
where ${\rm Cov}(\sv_k) = \Qm_k$ and $\zv \sim \Cc\Nc(\zerov, \Sigmam(\lambdav))$.
In this ``dual MAC'' interpretation, group sum-rate expression (\ref{eq:min-max-1}) corresponds to
group-wise successive interference cancellation, where user groups are decoded successively in the
order of $\pi_1, \pi_2, \ldots, \pi_A$, and users in each group are jointly decoded. Also, notice
that users in group $\pi_k$ in general do not achieve individually the rate $R_{\pi_k}(\widetilde{\Hm})$
on every slot. Rather, this rate is the aggregate sum-rate of all users in group $\pi_k$, normalized
by $N$, i.e., the mean user rate of group $\pi_k$ for given $\widetilde{\Hm}$.

Efficient interior-point methods to solve (\ref{eq:min-max}) are given, for example, in
\cite{Yu-TIT06, Zhang-Poor-etal-ISIT09, Huh-Caire-submitTSP09}. Yet, the solution of this problem is
numerically fairly involved, especially for large dimensions.

Consistently with the assumption of fixed coefficients $\{\beta_{m,k}\}$ and ergodic block-fading
for the small-scale fading coefficients $\{\Hm_{m,k}\}$, the {\em ergodic capacity region} of the
MU-MIMO downlink channel (\ref{eq:single-bc}) is given by the set of all achievable {\em average} rates,
where averaging is with respect to the small-scale fading coefficients. In particular, let
$R_k(\widetilde{\Hm}, W_1, \ldots, W_A)$ denote the $k$-th user group rate at the solution of
(\ref{eq:min-max}). Then, an inner bound to the ergodic capacity region is given by
\begin{align} \label{eq:c-erg}
\underline{\Cc}(P_1,\ldots, P_B) = {\rm coh} \bigcup_{W_1, \ldots, W_A \geq 0}
  & \left \{ \Rm: 0 \leq R_{k,i} \leq \EE \left[ R_k(\widetilde{\Hm}, W_1, \ldots, W_A) \right],
  \right. \nonumber \\
& \left. \;\; \forall k=1,\ldots,A, \; \forall i=1,\ldots,N \right\}
\end{align}
where ``coh'' indicates the closure of the convex hull. The achievability of the above region is
clear: all users $i$ in group $k$ are statistically equivalent and therefore they can achieve the
same ergodic rate. Notice that $\underline{\Cc}(P_1,\ldots, P_B)$ is generally an inner bound
because of the restriction of the weights in (\ref{eq:min-max}) to be identical for all users in
the same group. We will see later that, for fairness scheduling in the limit of
$N \rightarrow \infty$, this limitation becomes immaterial.

At this point we can formulate the fairness scheduling problem. Let $g(\Rm)$ denote a strictly
increasing and concave network utility of the ergodic user rates. While the channel fading
coefficients change from time slot to time slot according to some ergodic process, the optimal
scheduling policy allocates dynamically the transmit powers and the DPC precoding order in order
to let the system operate at the ergodic rate point solution of:
\begin{align} \label{eq:sche}
\mbox{maximize} \;\; & g(\Rm) \nonumber \\
\mbox{subject to} \;\; & \Rm \in \underline{\Cc}(P_1,\ldots, P_B)
\end{align}
Different fairness criteria can be enforced by choosing appropriately the function $g(\cdot)$
\cite{Mo-TNET00}. For example, proportional fairness \cite{Viswanath-Tse-Laroia-TIT02,
Bender-Viterbi-etal-CommMag00, Parkvall-Englund-Lundevall-Torsner-CommMag06} is obtained by
letting $g(\Rm) = \sum_{k,i} \log R_{k,i}$ and max-min fairness is obtained by letting
$g(\Rm) = \min_{k,i} R_{k,i}$.

We notice here that an analytical characterization of the ergodic rate point $\Rm^\star$ achieving
the optimum in (\ref{eq:sche}) is in general extremely complicated. However, by applying the general
stochastic optimization framework of \cite{Georgiadis-Neely-Tassiulas-04} (see also
\cite{ShiraniMehr-Caire-Neely-submit09}, more specifically targeted to the MU-MIMO downlink),
explicit scheduling policies can be designed
such that the limit of the time-averaged user rates converges to $\Rm^\star$.
The rest of this paper is dedicated to finding an efficient method to directly
compute $\Rm^\star$, by exploiting large random matrix theory and convex optimization.

\section{Weighted average sum rate maximization} \label{sec:wesrm}

In this section, we consider the solution of the following preliminary problem. For fixed
$\{\lambda_m\}$, we consider the maximization of the weighted average sum-rate maximization
\begin{align} \label{eq:wsrm}
\mbox{maximize} \;\; & \sum_{k=1}^A W_{\pi_k} \frac{1}{N} \EE \left[ \log
  \frac{\left| \Sigmam(\lambdav) + \widetilde{\Hm}_{k:A} \Qm_{k:A}
  \widetilde{\Hm}_{k:A}^\herm \right|}
  {\left| \Sigmam(\lambdav) + \widetilde{\Hm}_{k+1:A} \Qm_{k+1:A}
  \widetilde{\Hm}_{k+1:A}^\herm \right |} \right ]  \nonumber \\
\mbox{subject to} \;\; & \trace(\Qm) \leq Q
\end{align}
where we define $Q \eqdef \sum_{m=1}^B \lambda_m P_m$.
Letting $\Delta_k \eqdef W_{\pi_k} - W_{\pi_{k-1}}$ with $W_{\pi_0} = 0$, the objective function
in (\ref{eq:wsrm}) can be written as
\begin{equation} \label{eq:sum-rate-obj}
F_{\Wm,\lambdav}(\Qm) = \sum_{k=1}^A \Delta_k \frac{1}{N}\EE \left[ \log \left| \Sigmam(\lambdav) +
  \widetilde{\Hm}_{k:A} \Qm_{k:A} \widetilde{\Hm}_{k:A}^\herm \right| \right]
  - W_{\pi_A} \frac{1}{N} \log \left| \Sigmam(\lambdav) \right|
\end{equation}
In this form, problem (\ref{eq:wsrm}) is clearly convex, since $F_{\Wm,\lambdav}(\Qm)$ in
(\ref{eq:sum-rate-obj}) is a concave function of $\Qm$. In addition, we can prove the following
symmetry result:

\begin{lemma} \label{lemma:eqpow}
The optimal $\Qm$ in (\ref{eq:wsrm}) allocates equal power to the users in the same group.
\end{lemma}
\begin{IEEEproof}
Denote the utility function in (\ref{eq:sum-rate-obj}) as a function of diagonal entries of
$\Qm$ as $f(\{q_{k,i}: \; 1 \leq k \leq A , 1 \leq i \leq N, \})$. Since users in the same group
are statistically equivalent and the function $f(\cdot)$ is defined through an expectation with
respect to the channel fading coefficients, it follows that $f(\cdot)$ must be invariant with respect
to permutations of the arguments $q_{k,1}, \ldots, q_{k,N}$. That is, for any $k = 1,\ldots, A$,
and $1 \leq i < j \leq N$, the value of the function is invariant if the arguments $q_{k,i}$ and
$q_{k,j}$ are exchanged. Suppose that $\{q_{k,i}^*\}$ is the optimal input power allocation, solution
of (\ref{eq:wsrm}). Then, we have
\begin{align}
f(\ldots, q_{k,i}^*, \ldots, q_{k,j}^*, \ldots) &=
  f(\ldots, q_{k,j}^*, \ldots, q_{k,i}^*, \ldots) \nonumber \\
&\leq f(\ldots, \frac{q_{k,i}^*+q_{k,j}^*}{2}, \ldots, \frac{q_{k,i}^*+q_{k,j}^*}{2},
  \ldots). \nonumber
\end{align}
where the inequality follows from the concavity of $f(\cdot)$ and Jensen's inequality. Under the
optimality assumption, equality must hold and this implies that $\frac{q_{k,i}^*+q_{k,j}^*}{2}$
is the optimal input power for both users $i$ and $j$ in group $k$. Extending this argument by
induction, it follows that the optimal input power must be in the form $q_{k,i}^\star = Q_k/N$
for all $i = 1,\ldots, N$, for some values $Q_1, \ldots, Q_A$.
\end{IEEEproof}

Using Lemma \ref{lemma:eqpow}, we restrict the optimization in (\ref{eq:sum-rate-obj}) to
block-diagonal matrices $\Qm$ with constant diagonal blocks $\Qm_{k} = \frac{Q_k}{N}\Id$.
The following lemma shows that we can restrict to strictly positive $\{\lambda_m\}$:

\begin{lemma} \label{lemma:pos-lambda}
The optimal $\lambdav^\star$ for the min-max problem (\ref{eq:min-max}) are strictly positive,
i.e., $\lambdav^\star > \zerov$.
\end{lemma}
\begin{IEEEproof}
The dual variable $\lambda_m$
plays the role of the noise power at the antennas of the $m$-th BS in the dual MAC.
Let $G_\Wm(\lambdav) = \max_{\Qm:\trace(\Qm) \leq Q}  F_{\Wm,\lambdav}(\Qm)$ and suppose that
$\lambda_m^\star = 0$ for some $m$. Then, $|\Sigma(\lambdav^\star)|=0$ in (\ref{eq:sum-rate-obj}) and $G_\Wm(\lambdav^\star)$ goes to positive infinity, which clearly cannot be the solution of
the minimization with respect to  $\lambdav$ in (\ref{eq:min-max}).
Therefore, the optimal $\lambda_m^\star$ is strictly positive for all $m=1,\cdots,B$.
\end{IEEEproof}

Then we can define $\overline{\Hm}_{k} \eqdef \frac{1}{\sqrt{N}} \Sigmam^{-1/2} (\lambdav)
\widetilde{\Hm}_k$ and rewrite the objective function with some abuse of notation as
\begin{equation} \label{eq:obj-1}
F_{\Wm,\lambdav}(Q_1,\ldots, Q_A) = \sum_{k=1}^A \Delta_k \frac{1}{N} \EE \left[ \log \left| \Id +
  \sum_{\ell = k}^A \overline{\Hm}_{\pi_\ell} \overline{\Hm}_{\pi_\ell}^\herm Q_{\pi_\ell}
  \right| \right]
\end{equation}
where the trace constraint (\ref{eq:wsrm}) becomes $\sum_{k=1}^A Q_k \leq Q$.

\subsection{Solution for finite $N$}

The Lagrangian function of problem (\ref{eq:wsrm}) is given by
\begin{equation} \label{eq:lagrange-wesrm}
\Lc(Q_1,\ldots, Q_A; \xi) = F_{\Wm,\lambdav}(Q_1,\ldots, Q_A) - \xi \left ( \sum_{k=1}^A Q_k - Q \right )
\end{equation}
Using the differentiation rule $\partial \log |\Xm| = \trace ( \Xm^{-1} \partial \Xm)$, we write
the KKT conditions as
\begin{align} \label{eq:KKT-wesrm}
\frac{\partial \Lc}{\partial Q_{\pi_j}} \;=\; \sum_{k=1}^j \frac{\Delta_k}{N} \EE \left[ \trace
  \left( \overline{\Hm}_{\pi_j}^\herm \left[ \Id + \sum_{\ell=k}^A \overline{\Hm}_{\pi_\ell}
  \overline{\Hm}_{\pi_\ell}^\herm Q_{\pi_\ell} \right ]^{-1} \overline{\Hm}_{\pi_j}
  \right ) \right ] \leq \xi
\end{align}
for $j = 1,\ldots, A$, where equality must hold at the optimal point for all $j$ such that
$Q_{\pi_j} > 0$. After some algebra and the application of the Sherman-Morrison-Woodbury matrix
inversion lemma \cite{Horn-Johnson-85}, the trace in (\ref{eq:KKT-wesrm}) can be rewritten in a
more convenient form
\begin{align} \label{eq:mmse1}
\frac{1}{N} \trace \left( \overline{\Hm}_{\pi_j}^\herm \left[ \Id + \sum_{\ell=k}^A
  \overline{\Hm}_{\pi_\ell} \overline{\Hm}_{\pi_\ell}^\herm Q_{\pi_\ell} \right]^{-1}
  \overline{\Hm}_{\pi_j} \right) \;&=\;
  \frac{1}{N} \trace \left( \overline{\Hm}_{\pi_j}^\herm \Thetam_{k:A/j}^{-1} \overline{\Hm}_{\pi_j}
  \left[ \Id + Q_{\pi_j} \overline{\Hm}_{\pi_j}^\herm \Thetam_{k:A/j}^{-1} \overline{\Hm}_{\pi_j}
  \right ] ^{-1} \right ) \nonumber \\
&=\; \frac{1 - \MMSE_{k:A}^{(j)}}{Q_{\pi_j}}
\end{align}
where we let $\Thetam_{k:A/j} = \Id + \sum_{\ell=k, \ell\neq j}^A \overline{\Hm}_{\pi_\ell} \overline{\Hm}_{\pi_\ell}^\herm Q_{\pi_\ell}$ and where we define $\MMSE_{k:A}^{(j)}$ as follows:
consider the observation model
\begin{equation} \label{eq:sic-model}
\rv_{[k:A]} = \sum_{\ell=k}^A \sqrt{Q_{\pi_\ell}} \; \overline{\Hm}_{\pi_\ell} \sv_\ell + \zv
\end{equation}
where $\sv_k, \sv_{K+1}, \ldots, \sv_A$ and $\zv$ are Gaussian independent vectors with i.i.d.
components $\sim \Cc\Nc(0,1)$. Then, $\MMSE_{k:A}^{(j)}$ denotes the per-component MMSE for the
estimation of $\sv_j$ from $\rv_{[k:A]}$, for fixed (known) matrices $\overline{\Hm}_{\pi_k}, \ldots,
\overline{\Hm}_{\pi_A}$. Explicitly, we have
\begin{align} \label{eq:mmse}
\MMSE_{k:A}^{(j)} \;&=\; \frac{1}{N} \trace \left( \Id - Q_{\pi_j} \overline{\Hm}_{\pi_j}^\herm
  \left[ \overline{\Hm}_{\pi_j} \overline{\Hm}_{\pi_j}^\herm Q_{\pi_j} + \Thetam_{k:A/j} \right]^{-1}
  \overline{\Hm}_{\pi_j} \right) \nonumber \\
&=\; \frac{1}{N} \trace \left( \left[ \Id + Q_{\pi_j} \overline{\Hm}_{\pi_j}^\herm
  \Thetam_{k:A/j}^{-1} \overline{\Hm}_{\pi_j} \right] ^{-1} \right)
\end{align}
Using (\ref{eq:mmse1}) in (\ref{eq:KKT-wesrm}) and solving for the Lagrange multiplier, we find
\begin{equation} \label{eq:lagrange-wesrm-sol}
\xi =  \frac{1}{Q} \sum_{\ell=1}^A \sum_{k=1}^\ell \Delta_k (1 - \EE[\MMSE_{k:A}^{(\ell)}])
\end{equation}
Finally, we arrive at the conditions
\begin{equation} \label{KKT1}
Q_{\pi_j} = Q \frac{\sum_{k=1}^j \Delta_k (1 - \EE[\MMSE_{k:A}^{(j)}])}
{\sum_{\ell=1}^A \sum_{k=1}^\ell \Delta_k (1 - \EE[\MMSE_{k:A}^{(\ell)}])}
\end{equation}
for all $j$ such that $Q_{\pi_j} > 0$ where, using the KKT conditions and
(\ref{eq:lagrange-wesrm-sol}), we find that for all $j$ such that $Q_{\pi_j} = 0$, the inequality
\begin{equation} \label{eq:condition-pos}
Q \sum_{k=1}^j \frac{\Delta_k}{N} \EE \left [ \trace \left ( \overline{\Hm}_{\pi_j}^\herm
\Thetam_{k:A/j}^{-1} \overline{\Hm}_{\pi_j} \right ) \right ]
\leq \sum_{\ell=1}^A \sum_{k=1}^\ell \Delta_k (1 - \EE[\MMSE_{k:A}^{(\ell)}])
\end{equation}
must hold. Eventually, we have proved the following result:

\begin{theorem} \label{thm:finite-dim}
The solution $Q_1^\star, \ldots, Q_A^\star$ of problem (\ref{eq:wsrm}) is given as follows.
For all $j$ for which  (\ref{eq:condition-pos}) is satisfied, then $Q_{\pi_j}^\star = 0$.
Otherwise, the positive $Q_{\pi_j}^\star$ satisfy (\ref{KKT1}).
\hfill \IEEEQED
\end{theorem}

In finite dimension, an iterative algorithm that provably converges to the solution can be obtained
as a simple modification of \cite[Algorithm 1]{Tulino-TWC06}. The amount of calculation is tremendous
because the average MMSE terms must be computed by Monte Carlo simulation. Since our emphasis is on
the solution in the limit for $N \rightarrow \infty$, we omit these details and focus on the infinite
dimensional case in Section \ref{subsec:inf-wsrm}.
In addition, we have not yet addressed the outer minimization with respect to the Lagrange multipliers
$\{\lambda_m\}$. We postpone this issue to Section \ref{subsec:symm} where we discuss system symmetry
conditions for which the solution under the per-BS power constraint coincides with the laxer
per-cluster sum power constraint. In this case, we can let $\lambda_m = 1$ for all $m$, and no
minimization with respect to $\lambdav$ is needed.

\subsection{Limit for $N \rightarrow \infty$} \label{subsec:inf-wsrm}

In this section, we consider problem (\ref{eq:wsrm}) in the limit for $N \rightarrow \infty$, making
use of the asymptotic random matrix results of \cite{Tulino-TIT05}. In this regime, the instantaneous
per-user rates in (\ref{eq:min-max}) converge to their expected values by well-known convergence
results of the empirical distribution of the log-determinants in the rate expression
(\ref{eq:min-max-1}) \cite{Tulino-04, Girko-90}. Hence, in the large-system regime, the solution of
(\ref{eq:wsrm}) coincides with that of (\ref{eq:min-max}), for fixed channel pathloss coefficients
$\{\beta_{m,k}\}$, transmit power constraints $\{P_m\}$, weights $\{W_k\}$ and Lagrange multipliers
$\{\lambda_m\}$. We will use this fact in Section \ref{sec:fairness}, where we will examine a general
dynamic fairness scheduling policy for the actual (finite dimensional) system, and study its
performance in the large-system regime.

We introduce the normalized row and column indices $r$ and $t$, taking values in $[0,1)$, and the
aspect ratio of the matrix $\overline{\Hm}$ given by the ratio of the number of columns over the
number of rows and given by $\nu = \frac{A}{\gamma B}$. Then, we define the following piecewise
constant functions:
\begin{itemize}
\item $\Qc(t)$: (dual uplink) transmit power profile such that $\Qc(t) = Q_{\pi_k}$ for
    $\frac{k-1}{A} \leq t < \frac{k}{A}$.
\item $\Gc(r,t)$: channel gain profile of the matrix $\overline{\Hm}$ such that $\Gc(r,t) = \beta^2_{m,\pi_k}/\lambda_m$ for $\frac{m-1}{B} \leq r < \frac{m}{B}$ and $\frac{k-1}{A} \leq
    t < \frac{k}{A}$.
\item $\Upsilon_{k:A}(t)$: average per-component MMSE profile of the observation model
    (\ref{eq:sic-model}), such that $\Upsilon_{k:A}(t) = \MMSE_{k:A}^{(j)}$ for
    $\frac{k-1}{A} \leq t < 1$.
\item $\Gamma_{k:A}(t)$: signal-to-interference-plus-noise ratio (SINR) profile corresponding to
    $\Upsilon_{k:A}(t)$ such that $\Gamma_{k:A}(t) = 1/\Upsilon_{k:A}(t)-1$.
\end{itemize}
In the limit of large $N$, these functions satisfy equations given by the following lemma:

\begin{lemma} \label{lemma:fixedpt}
As $N \rightarrow \infty$, for each $k = 1,\ldots, A$, the SINR functions $\Gamma_{k:A}(t)$ satisfy
the fixed-point equation
\begin{equation} \label{eq:Gamma_t}
\Gamma_{k:A}(t) =  \int_0^1  \frac{\gamma B \Gc(r,t) \Qc(t) \; dr}
{1 + \nu \int_{(k-1)/A}^1  \frac{\gamma B \Gc(r,\tau) \Qc(\tau) \; d\tau}{1 + \Gamma_{k:A}(\tau)}}
\end{equation}
Also, the asymptotic $\Upsilon_{k:A}(t)$ is given in terms of the asymptotic $\Gamma_{k:A}(t)$ as
$\Upsilon_{k:A}(t)=1/(1+\Gamma_{k:A}(t))$.
\end{lemma}

\begin{IEEEproof}
We apply \cite[Lemma 1]{Tulino-TIT05} to the matrix
$\Id + \sum_{\ell=k}^A \overline{\Hm}_{\pi_\ell} \overline{\Hm}_{\pi_\ell}^\herm Q_{\pi_\ell}$
where $\overline{\Hm}_{k:A} = [\overline{\Hm}_{\pi_k}, \ldots, \overline{\Hm}_{\pi_A}]$
has independent non-identically distributed components. The variance profile in
\cite[Lemma 1]{Tulino-TIT05} is defined as the limit of the variance of the elements of the matrix
$\overline{\Hm}_{k:A}$, multiplied by the number of rows, $\gamma B N$. With our normalization,
the elements of each $(m,\ell)$-th block of $\overline{\Hm}_{k:A}$ of size $\gamma N \times N$
have variance $\frac{\beta_{m,\pi_\ell}^2/\lambda_m}{N}$. Therefore, the variance profile for the
application of \cite[Lemma 1]{Tulino-TIT05}  is given by $\gamma B \Gc(r,t)$, where $\Gc(r,t)$ is
the piecewise constant function defined above. Eventually, we arrive at (\ref{eq:Gamma_t}).
\end{IEEEproof}

Since all functions involved in Lemma \ref{lemma:fixedpt} are piecewise constant (although the
lemma applies in more generality), we can give a more explicit expression directly in terms of the
discrete set of values of these functions. Replacing $\Gamma_{k:A}(t) = \Gamma_{k:A}^{(j)}$ for
all $\frac{j-1}{A} \leq t < \frac{j}{A}$ with $j \geq k$ in (\ref{eq:Gamma_t}) and solving for
the integrals of piecewise constant functions, we obtain
\begin{align} \label{eq:fixedpt-sinr}
\Gamma_{k:A}^{(j)} & = \sum_{m = 1}^{B} \int_{\frac{m-1}{B}}^{\frac{m}{B}}
  \frac{\gamma B \Gc(r,t) \Qc(t) \;dr}{1 + \nu \sum_{\ell = k}^A
  \int_{\frac{\ell-1}{A}}^{\frac{\ell}{A}} \frac{\gamma B \Gc(r,\tau) \Qc(\tau) \;d\tau}
  {1 + \Gamma_{k:A}^{(\ell)}}} \nonumber \\
& = \gamma \sum_{m=1}^{B} \frac{(\beta_{m,{\pi_j}}^2/\lambda_m) Q_{\pi_j}}
  {1 + \sum_{\ell=k}^A \frac{(\beta_{m,\pi_\ell}^2/\lambda_m) Q_{\pi_\ell}}
  {1+\Gamma_{k:A}^{(\ell)}}}.
\end{align}
Combining (\ref{eq:fixedpt-sinr}) with the already mentioned modification of the iterative algorithm
of \cite[Algorithm 1]{Tulino-TWC06}, we obtain a procedure to compute the maximum weighted average
sum rate of problem (\ref{eq:wsrm}), for fixed weights $\{W_k\}$ and Lagrange multipliers
$\{\lambda_m\}$. This is summarized by Algorithm \ref{alg:wsrm} below (for notation simplicity,
the algorithm is written assuming $\pi_k = k$ for all $k = 1,\ldots, A$. We can always reduce to
this case after a simple reordering of the weights).

\begin{algorithm}
\caption{Input power optimization for weighted average sum rate maximization} \label{alg:wsrm}
\begin{enumerate}
\item Initialize $Q_k(0) = Q/A$ for all $k = 1,\ldots, A$.
\item For $i = 0,1,2,\ldots$, iterate until the following solution settles:
\begin{equation} \label{eq:alg1}
Q_{j}(i+1) = Q  \frac{\sum_{k=1}^j \Delta_{k} (1 - \Upsilon^{(j)}_{k:A}(i))}
    {\sum_{\ell=1}^A \sum_{k=1}^\ell \Delta_{k} ( 1 - \Upsilon^{(\ell)}_{k:A}(i))},
\end{equation}
for $j = 1, \ldots, A$, where $\Upsilon^{(j)}_{k:A}(i) = 1/(1 + \Gamma^{(j)}_{k:A}(i))$,
and $\Gamma^{(j)}_{k:A}(i)$ is obtained as the solution of the system of fixed point equations (\ref{eq:fixedpt-sinr}), also obtained by iteration, for powers $Q_{k} = Q_{k}(i), \; \forall k$.
\item Denote by $\Gamma^{(j)}_{k:A}(\infty)$, $\Upsilon^{(j)}_{k:A}(\infty)$ and by $Q_j(\infty)$
    the fixed points reached by the iteration at step 2). If the condition
\begin{equation} \label{eq:alg2}
Q \sum_{k=1}^j \Delta_{k} \Gamma^{(j)}_{k:A}(\infty) \leq \sum_{\ell=1}^A \sum_{k=1}^\ell \Delta_{k}
  \left( 1 - \Upsilon^{(\ell)}_{k:A}(\infty) \right)
\end{equation}
is satisfied for all $j$ such that $Q_j(\infty) = 0$, then stop. Otherwise, go back to the
initialization step, set $Q_j(0) = 0$ for $j$ corresponding to the lowest value of
$\sum_{k=1}^j \Delta_k \Gamma^{(j)}_{k:A}(\infty)$, and repeat steps 2) and 3) starting from the
new initial condition.
\end{enumerate}
\end{algorithm}

\subsection{Computation of the asymptotic rates} \label{subsec:asymp-rate}

After the powers $Q_k^\star = Q_k(\infty)$ have been obtained from Algorithm \ref{alg:wsrm}, it
remains to compute the corresponding average per-user rates. The average rate of users in group $k$
is given by
\begin{equation} \label{eq:indiv-rate}
R_{\pi_k} = \frac{1}{N}\EE \left[ \log \left| \Id + \sum_{\ell=k}^A
  \overline{\Hm}_{\pi_\ell} \overline{\Hm}_{\pi_\ell}^\herm Q_{\pi_\ell} \right| \right]
  - \frac{1}{N} \EE \left[ \log \left| \Id + \sum_{\ell=k+1}^A \overline{\Hm}_{\pi_\ell}
  \overline{\Hm}_{\pi_\ell}^\herm Q_{\pi_\ell} \right| \right]
\end{equation}
In the limit for $N\rightarrow \infty$,  we can use the asymptotic analytical expression for the
mutual information given in \cite{Aktas-TIT06}. Adapting \cite[Result 1]{Aktas-TIT06} to our case,
we obtain
\begin{align} \label{eq:sr-evans}
\lim_{N \rightarrow \infty} \frac{1}{N}\EE \left[ \log \left| \Id + \sum_{\ell=k}^A
  \overline{\Hm}_{\pi_\ell} \overline{\Hm}_{\pi_\ell}^\herm Q_{\pi_\ell} \right| \right]
\;=\; & \sum_{\ell = k}^A \log\left( 1 + \gamma Q^\star_{\pi_\ell} \sum_{m=1}^{B}
  (\beta_{m,\pi_\ell}^2/\lambda_m) u_m \right) \nonumber \\
& +\; \gamma \sum_{m=1}^{B} \log \left( 1 + \sum_{\ell = k}^A
  (\beta_{m,\pi_\ell}^2/\lambda_m) Q^\star_{\pi_\ell}  v_\ell \right) \nonumber \\
& -\; \gamma \sum_{\ell = k}^A \sum_{m=1}^{B} (\beta_{m,\pi_\ell}^2/\lambda_m)
  Q^\star_{\pi_\ell} u_m v_{\ell}
\end{align}
where for each $k = 1,\ldots, A$, $\{u_m: m = 1,\ldots, B\}$ and $\{v_\ell : \ell = k,\ldots, A\}$
are the unique solutions to the system of fixed point equations
\begin{align} \label{eq:sr-fixedpt}
u_m & = \left( 1 + \sum_{\ell = k}^A Q^\star_{\pi_\ell} ( \beta_{m,\pi_\ell}^2/\lambda_m ) v_\ell
  \right)^{-1}, ~~ m = 1,\ldots,B, \nonumber \\
v_\ell &= \left( 1 + \gamma \sum_{m=1}^{B} Q^\star_{\pi_\ell} (\beta_{m,\pi_\ell}^2/\lambda_m) u_m
  \right)^{-1}, ~~ \ell =k,\ldots,A.
\end{align}
The proof follows from \cite{Aktas-TIT06} based on the Girko's theorem \cite{Girko-90} (see also
\cite{Tulino-04}). Although (\ref{eq:sr-evans}) is not in a closed form, $\{u_m\}$ and $\{v_\ell\}$
in (\ref{eq:sr-fixedpt}) can be solved by fixed point iterations with $A + B$ variables. These
converge very quickly to the solution to any desired degree of numerical accuracy.

\subsection{System symmetry} \label{subsec:symm}

So far we have considered the solution of the maximization in (\ref{eq:wsrm}) for fixed
$\{\lambda_m\}$. However, we are interested in the solution of (\ref{eq:min-max})
including the per-BS power constraint, that requires minimization with respect to $\{\lambda_m\}$.
In finite dimension  and for fixed channel matrix, the min-max
problem can be solved by the subgradient-based iterative method of \cite{Zhang-Poor-etal-ISIT09} or
the infeasible-start Newton method of \cite{Yu-Lan-TSP07, Huh-Caire-submitTSP09}.
A direct application of these algorithms to the large system limit requires
asymptotic expressions for the subgradient with respect to $\{\lambda_m\}$ or the {\em KKT matrix}, respectively.
These quantities contain the second order derivatives of the Lagrangian function
with respect to $\{Q_k\}$ and $\{\lambda_m\}$, which do not appear to be amenable for
easily computable asymptotic limits.

A general method for the minimization with respect to $\{\lambda_m\}$ can be obtained as follows.
Let  $G_\Wm(\lambdav)$ denote the solution of (\ref{eq:wsrm}). This is a convex function of $\lambdav$
and the minimizing $\lambdav^\star$ must have strictly positive components by
Lemma \ref{lemma:pos-lambda}. Therefore, at the optimal point we have
$\left. \frac{\partial G_\Wm}{\partial \lambda_m} \right|_{\lambdav=\lambdav^\star} = 0$ for all $m=1,\cdots,B$.
It follows that  the solution can be approached by gradient descent
iterations where the gradient can be estimated by numerical differentiation
\cite{Cheney-Kincaid-04}. Let $\epsilonv_m$ be a $\gamma B N$-length vector for which the elements
$(m-1)\gamma N + 1, \cdots, m \gamma N$ are $\epsilon$ for some $\epsilon > 0$
and the other elements are zero. Then the approximation for the partial derivative of $G_\Wm(\lambdav)$ with
respect to  $\lambda_m$ is given by $\frac{G_\Wm(\lambdav+\epsilonv_m) - G_\Wm(\lambdav-\epsilonv_m)}{2\epsilon}$
with $O(\epsilon^2)$ error term \cite{Cheney-Kincaid-04}. Both $G_\Wm(\lambdav+\epsilonv_m)$ and
$G_\Wm(\lambdav-\epsilonv_m)$ are computed by Algorithm \ref{alg:wsrm}.

From the above argument it follows that the general case where minimization with respect to
$\{\lambda_m\}$ is required does not present any conceptual difficulty beyond the fact that it may be numerically cumbersome.
Of course, a simple upper bound consists of relaxing the per-BS power constraint to a sum-power constraint
in the reference cluster.  Notice that the solution and the value of the objective function is invariant
to a common scaling of the Lagrange multipliers. Therefore, we can assume
$\frac{1}{B} \sum_{m=1}^B \lambda_m = 1$ without loss of generality.
Letting $\lambda_m = 1$ for all $m$ yields the laxer sum-power constraint
$\sum_k Q_k \leq \sum_m P_m \eqdef P_{\rm tot}$, where $P_{\rm tot}$ denotes the total transmit power
of the cluster.  This choice yields an upper-bound to the capacity region of the cluster (under the
constraint of treating ICI as noise) and therefore also provides an upper-bound to the whole system
achievable region under the assumption that all BSs transmit at their maximum power.

\begin{figure}
\centering
\includegraphics[width=6in]{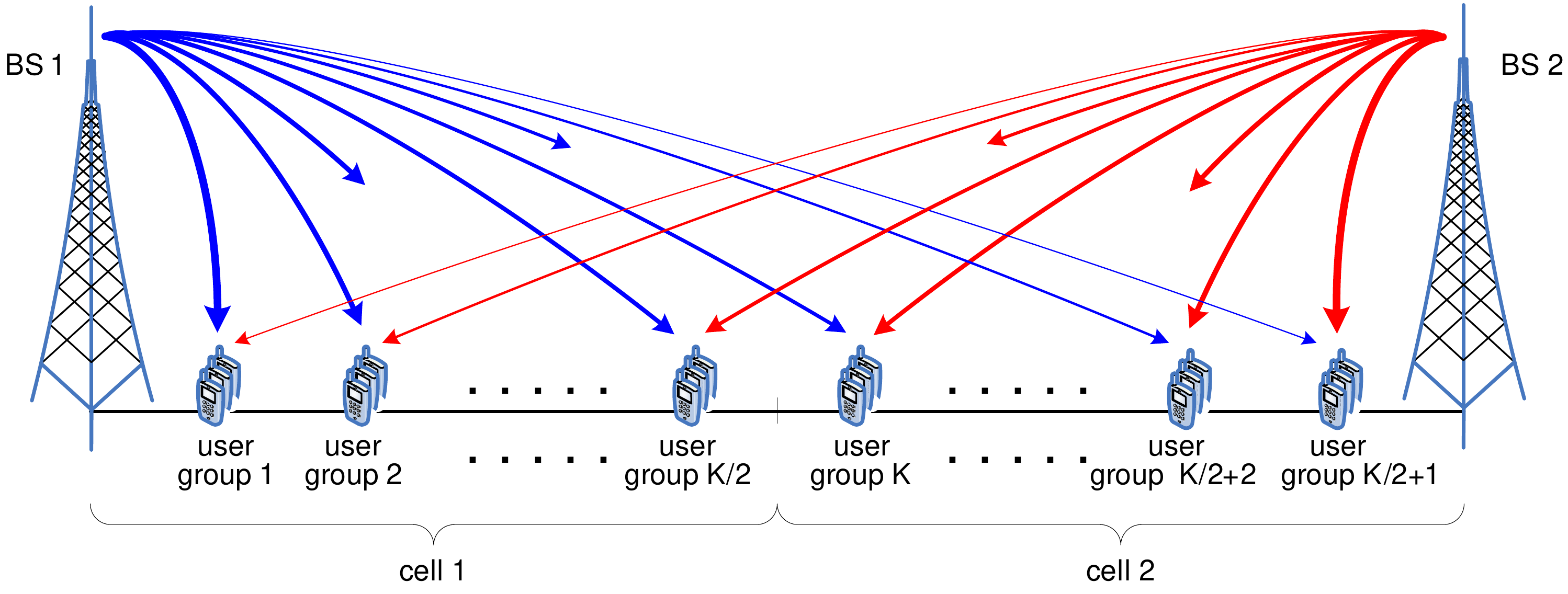}
\caption{A linear cellular layout with two cells and $K$ symmetric user groups.}
\label{fig:2cell-model}
\end{figure}

Next, we present a system symmetry condition under which the sum-power and the per-BS power solutions
coincide. Assume the same BS power constraint $P_m = P$ for all $m = 1,\ldots, B$. Then, let $A' = A/B$
assuming that $B$ divides $A$. In particular, this is true when we have the same number of user groups
in each cell of the cluster. Finally, assume that the $B \times A$ matrix of the coefficients
$\betav = \{\beta_{m,k}\}$ can be partitioned into $A'$ submatrices of size $B \times B$ such that
each submatrix has the property that all rows are permutations of the first row, and all columns are
permutations of the first column. Since this requirement reminds the condition for strongly symmetric
discrete memoryless channels, we shall refer to these submatrices as  ``strongly symmetric blocks''.
To fix ideas, consider Fig.~\ref{fig:2cell-model} showing a linear cellular layout with 2 cells and
$K$ user groups. Let $K = 8$ and assume distance-dependent pathloss coefficients yielding the matrix
\[ \betav = \left [ \begin{array}{cccccccc}
a & b & b & a & f & e & d & c \\
f & e & d & c & a & b & b & a
\end{array} \right ] \]
for some positive numbers $a,b,c,d,e,f$. We notice that this matrix can be decomposed into the
${A'} = 4$ strongly symmetric blocks
\[
\left[ \begin{array}{cc} a & f \\ f & a \end{array} \right], \;\;\;
\left[ \begin{array}{cc} b & e \\ e & b \end{array} \right], \;\;\;
\left[ \begin{array}{cc} b & d \\ d & b \end{array} \right], \;\;\;
\left[ \begin{array}{cc} a & c \\ c & a \end{array} \right]
\]
satisfying the above assumption.

When these symmetry condition hold, the user groups corresponding to the same strongly symmetric block
(e.g., user groups pairs $(1,5)$, $(2,6)$, $(3,7)$ and $(4,8)$ in the example) are statistically
equivalent, in the sense that they see exactly the same landscape of channel coefficients from all the
BSs forming the cluster. In this case, as it will be clear in Section \ref{sec:fairness}, we can restrict
the weighted sum-rate maximization in (\ref{eq:min-max}), (\ref{eq:wsrm}) to the case where the weights
$W_k$ are identical for all user groups in the same strongly symmetric block. Without loss of generality,
let's enumerate the user groups such that the $b$-th symmetric block contains user groups with indices
$k = (b-1)B + 1, \ldots, bB$, with corresponding constant weights $W_k  = W'_b$.
Then, the objective function (\ref{eq:obj-1}) takes on the form:
\begin{equation} \label{eq:obj-2}
F_{\Wm,\lambdav}(Q_1,\ldots, Q_A) = \sum_{b=1}^{A'} \frac{\Delta'_{b}}{N} \EE \left[ \log \left| \Id +
  \Sigmam^{-1}(\lambdav) \widetilde{\Hm}_{(b-1)B+1:A} \Qm_{(b-1)B+1:A}
  \widetilde{\Hm}_{(b-1)B+1:A}^\herm \right| \right]
\end{equation}
where $\Delta'_{b} = W'_{b+1} - W'_b$ for $b = 1,\ldots, {A'}$, with $W'_0 = 0$, and where
\[ \Qm = \frac{1}{N} \diag(\underbrace{Q_1, \ldots, Q_1}_{N}, \underbrace{Q_2, \ldots, Q_2}_{N},
  \ldots, \underbrace{Q_A, \ldots, Q_A}_{N}), \]
with trace constraint $\sum_{k=1}^A Q_k \leq BP = P_{\rm tot}$. We have the following result:

\begin{theorem} \label{thm:syst-symmetry}
Under the above system symmetry conditions, the minimization in the min-max problem (\ref{eq:min-max})
in the limit of $N \rightarrow \infty$ is achieved for $\lambda_m = 1$ for all $m = 1,\ldots, B$.
\end{theorem}

\begin{IEEEproof}
Let $\Psf_{i,j} = \EE \left[ \left| [\widetilde{\Hm}]_{i,j} \right|^2 \right]$ denote the variance
of the $(i,j)$-th element of the channel matrix. For any $b = 1,\ldots, {A'}$, the matrix
$\widetilde{\Hm}_{(b-1)B+1:A}$ has the property that the empirical distribution of the element
variances for all rows, i.e., the cumulative distribution functions
\[ \Fc^{(N)}_{i,b}(z) \eqdef \frac{1}{(A - (b-1)B)N} \sum_{j=(b-1)BN+1}^{AN}
  1 \left\{ \Psf_{i,j} \leq z \right\} \]
are the same, for all row index $i = 1, \ldots, \gamma BN$. This means that the matrix of the element
variances $\{\Psf_{i,j}\}$ corresponding to $\widetilde{\Hm}_{(b-1)B+1:A}$ is {\em row-regular}
(see definition in \cite[Definition 5]{Tulino-TIT05}). Under the row-regularity condition, it follows
that $\widetilde{\Hm}_{(b-1)B+1:A}$ in the limit of $N \rightarrow \infty$ is statistically equivalent
to a matrix $\check{\Hm}_{(b-1)B+1:A} = \Gm_{(b-1)B+1:A} \Tm^{1/2}_{(b-1)B+1:A}$, where
$\Gm_{(b-1)B+1:A}$ is an i.i.d. matrix with zero-mean, unit-variance elements, and $\Tm_{(b-1)B+1:A}$
is a non-negative diagonal matrix with asymptotic empirical spectral distribution given by
$\lim_{N\rightarrow \infty} \Fc^{(N)}_{i,b}(z)$. In particular, the distribution of
$\check{\Hm}_{(b-1)B+1:A}$ is asymptotically unitary left-invariant,  that is, for any unitary matrix
$\Um$ independent of $\check{\Hm}_{(b-1)B+1:A}$, the matrices $\Um \check{\Hm}_{(b-1)B+1:A}$ and
$\check{\Hm}_{(b-1)B+1:A}$ are asymptotically identically distributed.

Let $\Pim$ denote a $\gamma B N \times \gamma B N$ block-permutation matrix, that permutes the $B$
blocks of consecutive positions of length $\gamma N$ in the index vector $\{1, \ldots, \gamma B N\}$.
Using the above asymptotic statistical equivalence, in the limit of large $N$ we can write, for any
$\{\lambda_m\}$ and $\{Q_k\}$,
\begin{align} \label{eq:obj-3}
F_{\Wm,\lambdav}(Q_1,\ldots, Q_A) \; & \stackrel{(a)}{=} \;
  \sum_{b=1}^{A'} \frac{\Delta'_{b}}{N} \EE \left[ \frac{1}{B!} \sum_{\Pim} \log \left| \Id +
  \Sigmam^{-1}(\lambdav) \Pim \check{\Hm}_{(b-1)B+1:A} \Qm_{(b-1)B+1:A} \check{\Hm}_{(b-1)B+1:A}^\herm
  \Pim^\transp  \right| \right] \nonumber \\
& = \; \sum_{b=1}^{A'} \frac{\Delta'_{b}}{N} \EE \left[ \frac{1}{B!} \sum_{\Pim} \log \left | \Id +
  \Pim^\transp \Sigmam^{-1}(\lambdav) \Pim \check{\Hm}_{(b-1)B+1:A} \Qm_{(b-1)B+1:A}
  \check{\Hm}_{(b-1)B+1:A}^\herm \right| \right] \nonumber \\
& = \; \sum_{b=1}^{A'} \frac{\Delta'_{b}}{N} \EE \left[ \frac{1}{B!} \sum_{\Pim} \log \left | \Id +
  \left( \Pim^\transp \Sigmam(\lambdav) \Pim \right)^{-1} \check{\Hm}_{(b-1)B+1:A} \Qm_{(b-1)B+1:A}
  \check{\Hm}_{(b-1)B+1:A}^\herm \right| \right] \nonumber \\
& \stackrel{(b)}{\geq} \; \sum_{b=1}^{A'} \frac{\Delta'_{b}}{N} \EE \left[ \log \left| \Id +
  \left( \frac{1}{B!} \sum_{\Pim}  \Pim^\transp \Sigmam(\lambdav) \Pim \right)^{-1}
  \check{\Hm}_{(b-1)B+1:A} \Qm_{(b-1)B+1:A} \check{\Hm}_{(b-1)B+1:A}^\herm \right| \right] \nonumber \\
& \stackrel{(c)}{=} \; \sum_{b=1}^{A'} \frac{\Delta'_{b}}{N} \EE \left[ \log \left| \Id +
  \check{\Hm}_{(b-1)B+1:A} \Qm_{(b-1)B+1:A} \check{\Hm}_{(b-1)B+1:A}^\herm \right| \right]
\end{align}
where (a) follows from the left-unitary invariance, (b) follows from Jensen's inequality and (c) from
the fact that, without loss of generality, we let $\frac{1}{B} \sum_{m=1}^B \lambda_m = 1$.
This shows that, for asymptotically large $N$ and under the given symmetry conditions of the channel
coefficients and rate weights, the worst-case Lagrange multipliers for the weighted maximization of
the average rates in (\ref{eq:wsrm}) is $\lambda_m = 1$. Since for $N \rightarrow \infty$ the
instantaneous rates in (\ref{eq:min-max}) converge to the average rates in (\ref{eq:wsrm}), the theorem
is proved.
\end{IEEEproof}

\section{Fairness scheduling} \label{sec:fairness}

Downlink opportunistic scheduling is currently used by ``high data rate'' third-generation cellular
systems such as EV-DO \cite{Bender-Viterbi-etal-CommMag00} and HSDPA
\cite{Parkvall-Englund-Lundevall-Torsner-CommMag06}. It is expected that in the next generation of
systems based on MIMO-OFDM, such as IEEE 802.16m \cite{IEEE80216m} and LTE-Advanced \cite{LTE-Advanced},
such strategies will be integrated with the MU-MIMO physical layer.
In such systems, each cooperation cluster runs a downlink scheduler that computes a set of rate weight
coefficients and, at each scheduling time slot $t$, solves the maximization of the instantaneous
weighted rate-sum subject to the per-BS power constraint, as in (\ref{eq:min-max}). The result of this
maximization provides the power and rate allocation and the corresponding downlink precoder parameters
(i.e., the beamforming vectors and the DPC encoding order) to be used in the current slot.
The scheduler weights are recursively computed such that the time-averaged user rates converge to the
desired ergodic rate point $\Rm^\star$, the solution of (\ref{eq:sche}).

The scheduling policy can be systematically designed by using the stochastic optimization approach of
\cite{Georgiadis-Neely-Tassiulas-04, ShiraniMehr-Caire-Neely-submit09}, based on the idea of ``virtual
queues''. Notice that we do not consider {\em exogenous} arrivals:
consistently with the classical information theoretic setting, we assume that an arbitrarily large number
of information bits are to be transmitted to the users in each cluster (infinitely backlogged system).
The virtual queues defined here are only a tool to recursively compute the weights of the scheduling
policy. In order to illustrate the scheduling mechanism we will denote {\em instantaneous} quantities
as dependent on the slot index $t$. In short, the policy ensures that the virtual queue of each user
$(k,i)$ (i.e., user $i$ in group $k$) is {\em strongly stable} (see
\cite[Definition 3.1]{Georgiadis-Neely-Tassiulas-04}). This implies that the arrival rate
$\Lambda_{k,i}$ is strictly less than the average service rate
$R_{k,i} = \EE[R_{k,i}(\widetilde{\Hm}(t))]$. Then, the desired ergodic rate point $\Rm^\star$ can be
approached if the virtual queues are fed by virtual arrival processes $A_{k,i}(t)$ with arrival rates
$\Lambda_{k,i} = \EE[A_{k,i}(t)]$ sufficiently close to the desired values $R_{k,i}^\star$.
The interesting feature of this approach is that it is possible to generate such virtual arrival
processes adaptively, even if the values $R_{k,i}^\star$ are unknown a priori, and may be very difficult
to be calculated directly.

Let $U_{k,i}(t)$ denote the virtual queue backlog for user $i$ in group $k$ at time slot $t$, evolving
according to the stochastic difference equation
\begin{equation} \label{eq:queue}
U_{k,i}(t+1) = \left [ U_{k,i}(t)-R_{k,i}(\widetilde{\Hm}(t)) \right ]_+ + A_{k,i}(t)
\end{equation}
We consider the scheduling policy given as follows:
\begin{itemize}
\item At each time slot $t$, solve the weighted sum-rate maximization problem
\begin{align} \label{eq:dynwsrm}
\mbox{maximize} & \;\; \sum_{k=1}^A \sum_{i=1}^N U_{k,i}(t) R_{k,i}(\widetilde{\Hm}(t)) \nonumber \\
\mbox{subject to} & \;\; {\rm Cov}(\xv_m) \leq P_m
\end{align}
\item The virtual queues are updated according to (\ref{eq:queue}), where the arrival processes are
given by $A_{k,i}(t) = a_{k,i}^\star$, where the vector $\av^\star$ is the solution of  the maximization
problem:
\begin{align} \label{eq:dynfc}
\mbox{maximize} & \;\; V g(\av) - \sum_{k=1}^A \sum_{i=1}^N a_{k,i} U_{k,i}(t) \nonumber \\
\mbox{subject to} & \;\; 0 \leq a_{k,i} \leq A_{\max}
\end{align}
for suitable $V>0$ and $A_{\max}>0$.
\end{itemize}
The parameters $V$ and $A_{\max}$ determine the accuracy and the rate of convergence of the time-average
rates to their expected values. It can be shown \cite{Georgiadis-Neely-Tassiulas-04,
ShiraniMehr-Caire-Neely-submit09} that, for fixed sufficiently large parameters $A_{\max}$, the gap
between the long-time average rates
$\lim_{t \rightarrow \infty} \sum_{\tau=0}^{t-1} \frac{1}{t} R_{k,i}(\widetilde{\Hm}(\tau))$ and the
optimal ergodic rates $R_{k,i}^\star$ decrease as $O(1/V)$ while the expected backlog of the virtual
queues increases as $O(V)$.

After reviewing the above background on scheduling and stochastic optimization, we are ready to make some
observations that are instrumental for the performance computation in the large-system limit.
Due to the statistical equivalence of users in the same group, the ergodic rate points with $R_{k,i} = R_k$
(independent of $i$) are achievable. In particular, the boundary of the system ergodic capacity region
and of the region $\underline{\Cc}(P_1,\ldots, P_B)$ in (\ref{eq:c-erg}) coincide for all rate points
meeting this condition. It is meaningful to assume that the network utility function $g(\Rm)$ is
invariant with respect to permutations of the rates of statistically equivalent users. In fact, all
statistically equivalent users should be treated equally in the long-term average sense.\footnote{Here
we assume that all users have equal priority. For example, they are all delay-tolerant data users with
no particular individual priority: users differ only by their location in the cluster, which determines
their channel coefficients $\{\beta_{k,m}\}$.} For example, the $\alpha$-fairness utility function proposed
in \cite{Mo-TNET00} satisfies this condition. In this case, it is immediate to show that the function
$g(\Rm)$ is maximized by some rate point with equal rates over each user group or, if the symmetry
conditions of Theorem \ref{thm:syst-symmetry} hold, over all groups in the same strongly symmetric block.
Hence, in large-system limit the point $\Rm^\star = \{R_{k,i}^\star\}$ solution of (\ref{eq:sche})
must satisfy, for all $i$,
\[ R_{\pi_k,i}^\star = \lim_{N \rightarrow \infty}
  \frac{1}{N} \EE \left[ \log \frac{\left| \Id  + \sum_{\ell = k}^A \overline{\Hm}_{\pi_\ell}
  \overline{\Hm}_{\pi_\ell}^\herm Q_{\pi_\ell} \right|}
  {\left| \Id + \sum_{\ell = k+1}^A \overline{\Hm}_{\pi_\ell} \overline{\Hm}_{\pi_\ell}^\herm
  Q_{\pi_\ell} \right|} \right] \]
where the term on the right-hand side is the average per-user rate given by the solution of
(\ref{eq:wsrm}) for some choice of the weights $\{W_k\}$ and Lagrange multipliers $\{\lambda_m\}$.
It is well-known that, for a deterministic network, the dynamic scheduling policy described before
coincides with the Lagrangian dual optimization with outer subgradient iteration, where the Lagrangian
dual variables play the role of the virtual queues backlogs in the dynamic setting. In the large-system
limit, the channel uncertainty disappears and the MU-MIMO system ``freezes'' to a deterministic limit.
Using the large-system limit solution of (\ref{eq:wsrm}) presented in Section \ref{sec:wesrm}, the
solution of the fairness scheduling problem (\ref{eq:sche}) can be addressed directly, using Lagrangian
duality.

\subsection{Lagrangian optimization} \label{subsec:lagrange}

We rewrite (\ref{eq:sche}) using the auxiliary variables $\rv = [r_1, \ldots, r_A]$ and using the
definition of the ergodic rate region (\ref{eq:c-erg}) as:
\begin{align} \label{eq:aux}
\min_{\lambdav} \; \max_{\rv,\Qm,\pi} \;\;\; &  g(\rv) \nonumber \\
\mbox{subject to} \;\;\; &  r_{\pi_k} \leq \frac{1}{N} \EE \left[ \log
  \frac{\left| \Id  + \sum_{\ell = k}^A \overline{\Hm}_{\pi_\ell} \overline{\Hm}_{\pi_\ell}^\herm
  Q_{\pi_\ell} \right |}
  {\left| \Id + \sum_{\ell = k+1}^A \overline{\Hm}_{\pi_\ell} \overline{\Hm}_{\pi_\ell}^\herm
  Q_{\pi_\ell} \right |} \right ], \nonumber \\
& \trace(\Qm) \leq Q, \;\;\; \lambdav \geq 0
\end{align}
The Lagrange function for (\ref{eq:aux}) is given by
\begin{align} \label{eq:lagrange}
\Lc(\lambdav,\rv,\Qm,\pi,\Wm) &= g(\rv) - \sum_{k=1}^A W_{\pi_k} \left( r_{\pi_k} -
  \frac{1}{N} \EE \left[ \log \frac{\left| \Id  + \sum_{\ell = k}^A \overline{\Hm}_{\pi_\ell}
  \overline{\Hm}_{\pi_\ell}^\herm Q_{\pi_\ell} \right|}
  {\left| \Id + \sum_{\ell = k+1}^A \overline{\Hm}_{\pi_\ell} \overline{\Hm}_{\pi_\ell}^\herm
  Q_{\pi_\ell} \right|} \right]
  \right) \nonumber \\
&= \underbrace{g(\rv) - \sum_{k=1}^A W_k r_k }_{f_{\Wm}(\rv)} +
  \underbrace{\sum_{k=1}^A W_{\pi_k} \frac{1}{N} \EE \left[ \log
  \frac{\left| \Id  + \sum_{\ell = k}^A \overline{\Hm}_{\pi_\ell} \overline{\Hm}_{\pi_\ell}^\herm
  Q_{\pi_\ell} \right|}
  {\left|  \Id + \sum_{\ell = k+1}^A \overline{\Hm}_{\pi_\ell} \overline{\Hm}_{\pi_\ell}^\herm
  Q_{\pi_\ell} \right|} \right]
  }_{h_{\Wm}(\lambdav,\Qm,\pi)}
\end{align}
where $\Wm$ is the vector of dual variables corresponding to the auxiliary variable constraints
(rate constraints). The Lagrange function can be decomposed into a sum of a function of $\rv$ only,
denoted by $f_{\Wm}(\rv)$, and a function of $\lambdav,\Qm$ and $\pi$ only, denoted by $h_{\Wm}(\Qm,\pi)$.
The Lagrange dual function for the problem (\ref{eq:lagrange}) is given by
\begin{align} \label{eq:dual-ftn}
\Gc(\Wm) \;=& \; \min_{\lambdav} \max_{\rv,\Qm,\pi} \;\;\; \Lc(\lambdav,\rv,\Qm,\pi,\Wm) \nonumber \\
=& \; \underbrace{\max_{\rv} \;\;\; f_{\Wm}(\rv)}_{(a)}
  + \underbrace{\min_{\lambdav} \max_{\Qm,\pi} \;\;\; h_{\Wm}(\lambdav,\Qm,\pi)}_{(b)}
\end{align}
and it is obtained by the decoupled maximization in (a) (with respect to $\rv$) and  the min-max in (b)
(with respect to $\lambdav,\Qm, \pi$). Notice that problems (a) and (b) correspond to the static forms
of (\ref{eq:dynfc}) and (\ref{eq:dynwsrm}), respectively, after identifying $\rv$ with the virtual
arrival rates $\Am(t)$ and $\Wm$ with the virtual queue backlogs $\Um(t)$.
Finally, we can solve the dual problem defined as
\begin{equation} \label{eq:dual-prob}
\min_{\Wm \geq \zerov} \;\;\; \Gc(\Wm)
\end{equation}
Eventually, the solution of (\ref{eq:dual-prob}) can be found via inner-outer iterations as follows:

{\bf Inner Problem}: For given $\Wm$, we solve (\ref{eq:dual-ftn}) with respect to $\lambdav$,
$\rv$, $\Qm$ and $\pi$. This can be further decomposed into:
\begin{itemize}
\item Subproblem (a): Since $f_{\Wm}(\rv)$ is concave in $\rv \geq 0$, the optimal $\rv^*$ readily
    obtained by imposing the KKT conditions.
\item Subproblem (b): Taking the limit of $N \rightarrow \infty$, this problem is solved by
    Algorithm \ref{alg:wsrm} for fixed $\lambdav > 0$. If the system satisfies the symmetry conditions
    of Theorem \ref{thm:syst-symmetry} hold, then we let $\lambda_m = 1$ and no minimization with respect to $\lambda_m$ is needed. If these conditions do not hold, the outer minimization can be solved by the
    gradient descent method with the approximated gradient.
    Otherwise, letting $\lambda_m = 1$ yields an upper bound on the achievable network utility function,
    corresponding to the relaxation of the per-BS power constraint to the sum-power constraint.
\end{itemize}

{\bf Outer Problem}:
the minimization of $\Gc(\Wm)$ with respect to $\Wm \geq \zerov$ can be obtained by subgradient
adaptation. Let $\lambdav^*$, $\pi^*$, $\Qm^*$ and $\rv^*(\Wm)$ denote~\footnote{It is useful to
explicitly point out the dependence of $\Wm$ only for $\rv^*(\Wm)$, since this appears in the
subgradient expression, although it is clear that $\lambdav^*, \pi^*$ and $\Qm^*$ also in general
depend on $\Wm$.} the solution of the inner problem for fixed $\Wm$. For any $\Wm'$, we have
\begin{align} \label{eq:subgrad}
\Gc(\Wm') & =\; \max_{\rv} \; f_{\Wm'}(\rv) + \max_{\Qm} \; h_{\Wm'} (\lambdav^*,\Qm,\pi^*) \nonumber \\
&\geq f_{\Wm'}(\rv^*(\Wm)) + h_{\Wm'}(\lambdav^*,\Qm^*,\pi^*) \nonumber \\
&= \Gc(\Wm) + \sum_{k=1}^A \left( W'_{k} - W_{k} \right) \left( R_{k}^*(\Wm) - r_{k}^*(\Wm) \right)
\end{align}
where $R_{k}^*(\Wm)$ denotes the $k$-th group rate resulting from the solution of the inner problem
with weights $\Wm$, which is efficiently calculated  by Algorithm \ref{alg:wsrm} in the large-system
regime. A subgradient for $\Gc(\Wm)$ is given by the vector with components $R_k^*(\Wm) - r_k^*(\Wm)$.
Eventually, the dual variables $\Wm$ can be updated at the $n$-th outer iteration according to
\begin{align} \label{eq:muupdate}
W_k(n+1) &= W_k(n) - \mu(n) \left( R_{k}^*(\Wm(n)) - r_k^* (\Wm(n)) \right), \;\; \forall \; k
\end{align}
for some step size $\mu(n) > 0$ which can be determined by a efficient algorithm such as the
back-tracking line search method \cite{Boyd-Vandenberghe-04} or Ellipsoid method
\cite{Bertsimas-Tsitsiklis-97}. In the numerical example of Section \ref{sec:result}, we use the
back-tracking line search method. It should be noticed that by setting $\mu(n) = 1$ this subgradient
update plays the role of the virtual queue update in the dynamic scheduling policy of (\ref{eq:queue}).
But in this optimization, the objective function converges to a single optimal point by the iterations
and, by adjusting the step size $\mu(n)$ with the above methods, the convergence can be attained
very fast.

As an application example of the above general optimization, we focus on the two special cases of
{\em proportional fairness scheduling} (PFS) and {\em hard-fairness scheduling} (HFS), also known
as max-min fairness scheduling.

\subsection{Proportional fairness scheduling} \label{subsec:pfs}

The network utility function for PFS is given as
\begin{equation} \label{eq:pfs-util}
g(\rv) = \sum_{k=1}^A \log (r_k)
\end{equation}
In this case, the KKT conditions for the inner subproblem (a) yield the solution
\begin{align} \label{eq:sola-pfs}
r^*_k(\Wm) =  1/W_k, \;\; \forall \; k
\end{align}
(notice that $r_k$ must be positive for all $k$ otherwise the objective function is $-\infty$).
As mentioned before, the dual variables play the role of the virtual queue backlogs in the dynamic
scheduling policy, while the auxiliary variables $\rv$ correspond to the virtual arrival rates.
From (\ref{eq:sola-pfs}), we see that at the $n$-th outer iteration these variables are related by
$W_k(n) =\frac{1}{r^*_{k}(\Wm(n))}$. As observed at the beginning of Section \ref{sec:fairness},
the virtual arrival rates of the dynamic scheduling policy are designed in order to be close to the
ergodic rates $\Rm^\star$ at the optimal fairness point. It follows that the usual intuition of PFS,
according to which the scheduler weights are inversely proportional to the long-term average user
rates, is recovered.

\subsection{Hard fairness scheduling} \label{subsec:hfs}

In case of HFS, the scheduler maximizes the minimum user ergodic rate. The network utility function
is given by
\begin{equation} \label{eq:hfs-util}
g(\rv) =  \min_{k=1,\ldots,A}  r_k.
\end{equation}
This objective function is not strictly concave and differentiable everywhere. Therefore, it is
convenient to rewrite subproblem (a) by introducing an auxiliary variable $\gamma$, as follows:
\begin{align}
\max_{\gamma, \rv \geq 0} & \;\;\; \gamma - \sum_{k=1}^A W_k r_k  \nonumber \\
\mbox{subject to} & \;\;\; r_k \geq \gamma, \;\; \forall \; k
\end{align}
The solution must satisfy $r_k = \gamma$ for all $k$, leading to
\begin{equation} \label{eq:proba-hpf}
\max_{\gamma > 0} \;\;\; (1 - \sum_{k=1}^A W_k) \gamma.
\end{equation}
Since the original maximization in (\ref{eq:aux}) is bounded while (\ref{eq:proba-hpf}) may be
unbounded, we must have that $\sum_{k=1}^A W_k = 1$ and $\gamma$ must take on some appropriate value
that enforces this condition. The subgradient iteration for the weights $\Wm$, using
$r_k^*(\Wm(n)) = \gamma^*(\Wm(n))$, becomes
\begin{equation} \label{eq:muupdate-hfs}
W_k (n+1) = W_k(n) - \mu \left( R_{k}^*(\Wm(n)) - \gamma^*(\Wm(n)) \right), \;\; \forall \; k
\end{equation}
Summing up the update equations over $k=1,\ldots,A$ and using the conditions that
$\sum_{k=1}^A W_k(n) = 1$ for all $n$, we obtain
\begin{equation} \label{eq:hfsrate}
r_k^*(\Wm(n)) = \gamma^*(\Wm(n)) = \frac{1}{A} \sum_{j=1}^A R^*_{j}(\Wm(n)), \;\; \forall \; k
\end{equation}

\section{Numerical Results} \label{sec:result}


In this section we present some examples of the multi-cell model considered in this paper and compare
(when possible) the numerical results using the proposed large-system analysis with the results of
Monte Carlo simulation applied to an actual finite-dimensional system subject to the dynamic fairness
scheduling policy outlined at the beginning of Section \ref{sec:fairness}.

The examples involve a one-dimensional 2-cell model ($M=2$) and a two-dimensional three-sectored
7-cell model ($M=21$). In both cases, the system parameters and pathloss model are based on the mobile
WiMAX system evaluation specification \cite{Wimax-eval06} with cell radius 1.0 km and no shadowing
assumption. The 2-cell model, shown in Fig. \ref{fig:2cell-model}, considers two one-sided BSs with
$\gamma=4$, located at $\pm1$ km, and $K=8$ user groups equally spaced between the BSs.
We consider the case of full BS cooperation and no cooperation with a symmetric partition of users,
yielding $L=2$ clusters with $\Kc_1=\{1,2,3,4\}$ and $\Kc_2=\{5,6,7,8\}$.

\begin{figure}
\centering
\subfigure[Convergence of the utility function]{
\includegraphics[width=4.5in]{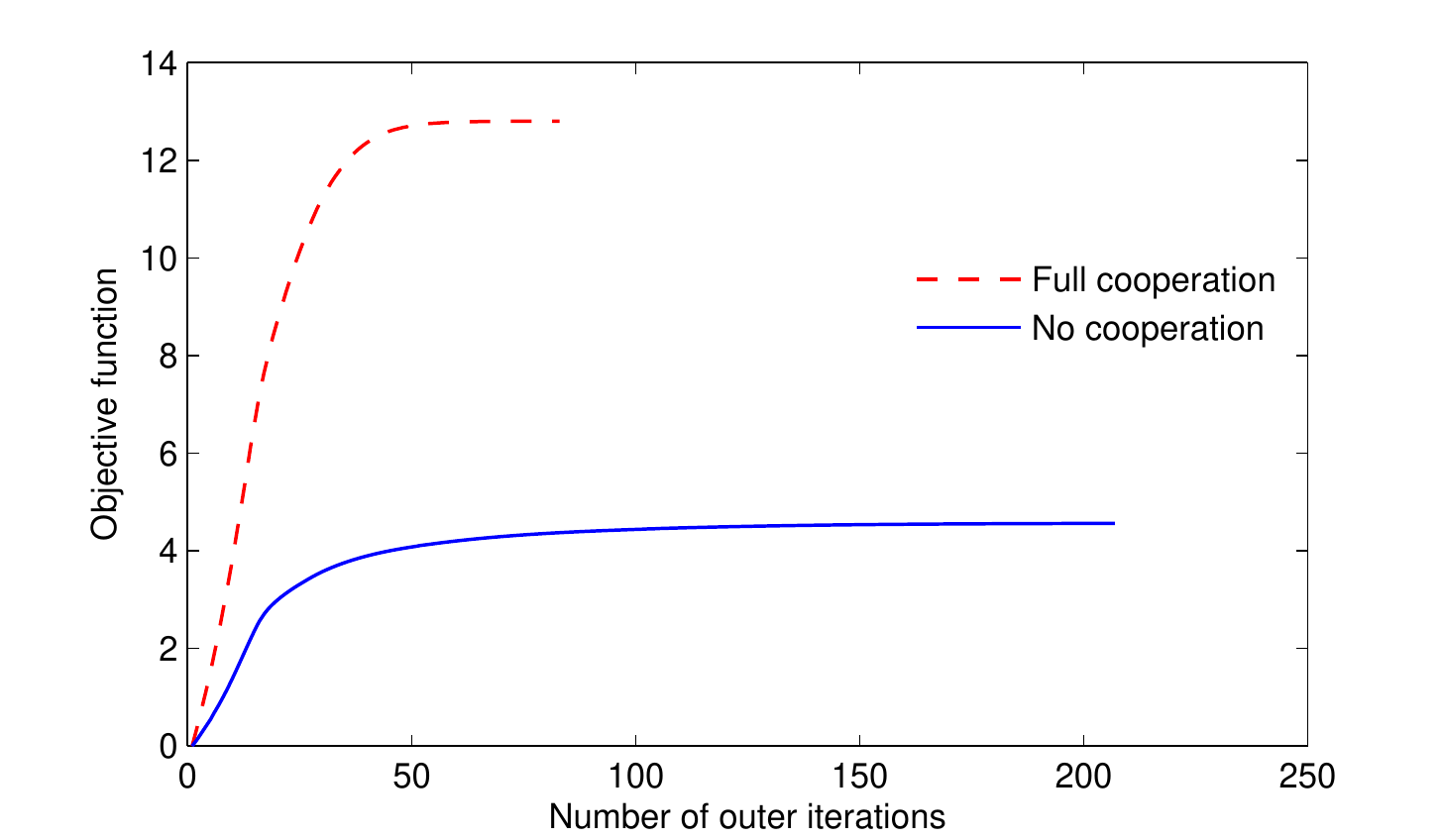}
\label{subfig:2cell-pfs-conv}
}
\subfigure[Individual group rates]{
\includegraphics[width=4.5in]{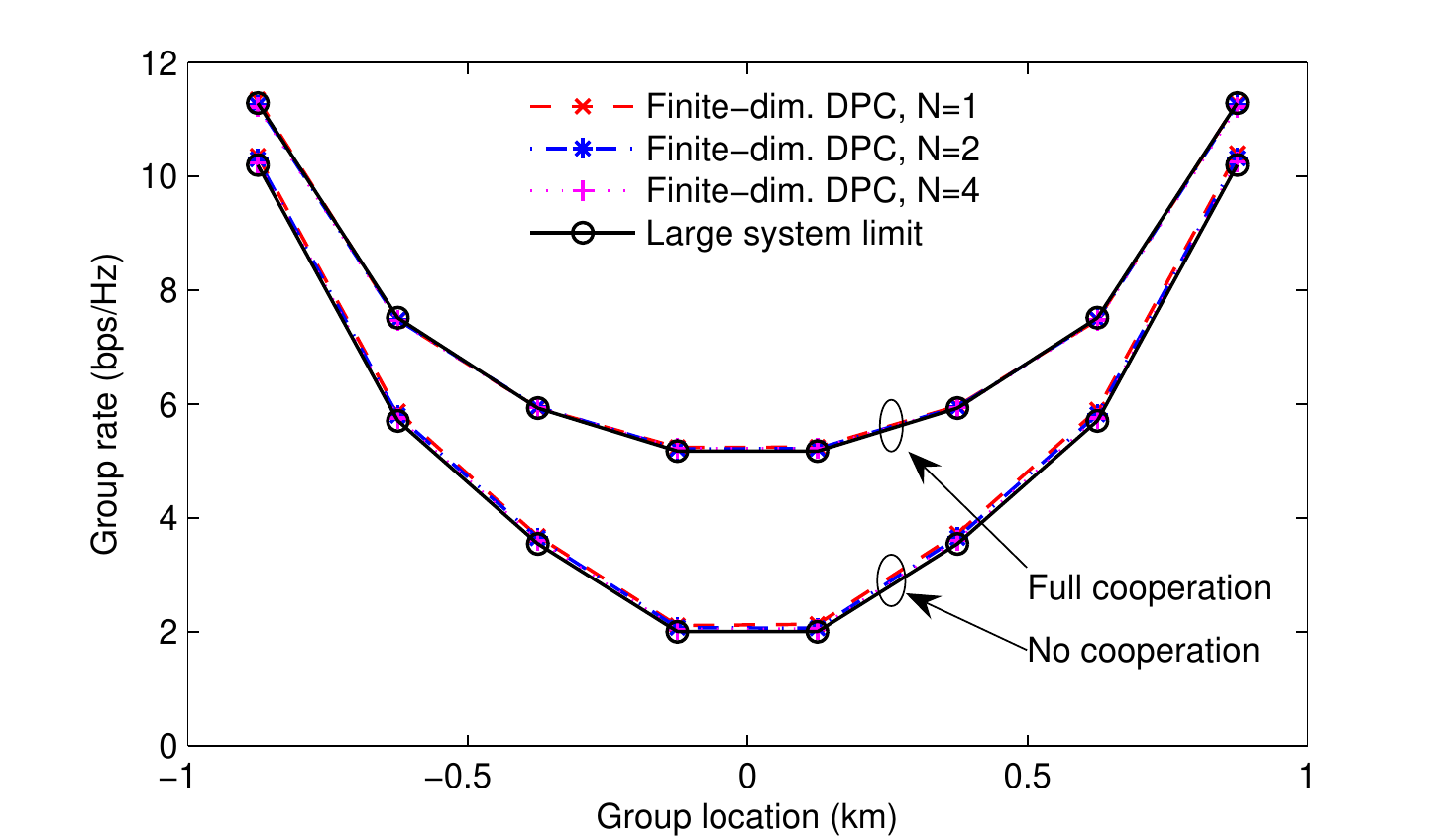}
\label{subfig:2cell-pfs-rate}
}
\caption{Proportional fairness scheduling with $\gamma=4$ and $K=8$ in the 2-cell model.}
\label{fig:2cell-pfs}
\end{figure}

Fig. \ref{fig:2cell-pfs} illustrates the convergence of the utility function and individual
group rates under PFS. In Fig. \ref{subfig:2cell-pfs-conv}, the PFS objective functions in the
no cooperation and full cooperation cases are shown to converge to the respective optimal PFS points.
Not surprisingly, the full cooperative system achieves significantly higher value of the objective
function (sum of the log-rates). In Fig. \ref{subfig:2cell-pfs-rate}, we compare the asymptotic rates
in the large-system limit with the achievable rates obtained by using Monte Carlo simulation in
finite dimension. In finite dimension we considered $N$ = 1, 2, or 4 and the same parameters of the
infinite-dimensional case. The channel vectors are randomly generated and change at every $t$ in an
i.i.d. fashion, and the instantaneous rates are allocated by using the DPC with the water-filling
algorithm \cite{Yu-TIT06} combined with the dynamic scheduling policy
\cite{ShiraniMehr-Caire-Neely-submit09} outlined in Section \ref{sec:fairness}.
Remarkably, the finite-dimensional simulation produced nearly the same rates for the considered
values of $N$ and these rates also almost overlap with the the large-system asymptotic results even
for very small $N$. Notice that the dynamic scheduling policy should provide multi-user diversity gain
and in general should achieve higher rates than the large-system limit, which is not able to exploit
the dynamic fluctuations of the small-scale fading due to ``channel hardening''. However, it appears
that in the regime where the pathloss is dominant over the randomness of the multi-antenna channels
and the number of users is not much larger than the number of BS antennas, the multi-user diversity
gain is negligible and the asymptotic analysis generates results very close to the simulations with
dynamic scheduling and DPC.

\begin{figure}
\centering
\subfigure[Convergence of the utility function]{
\includegraphics[width=4.5in]{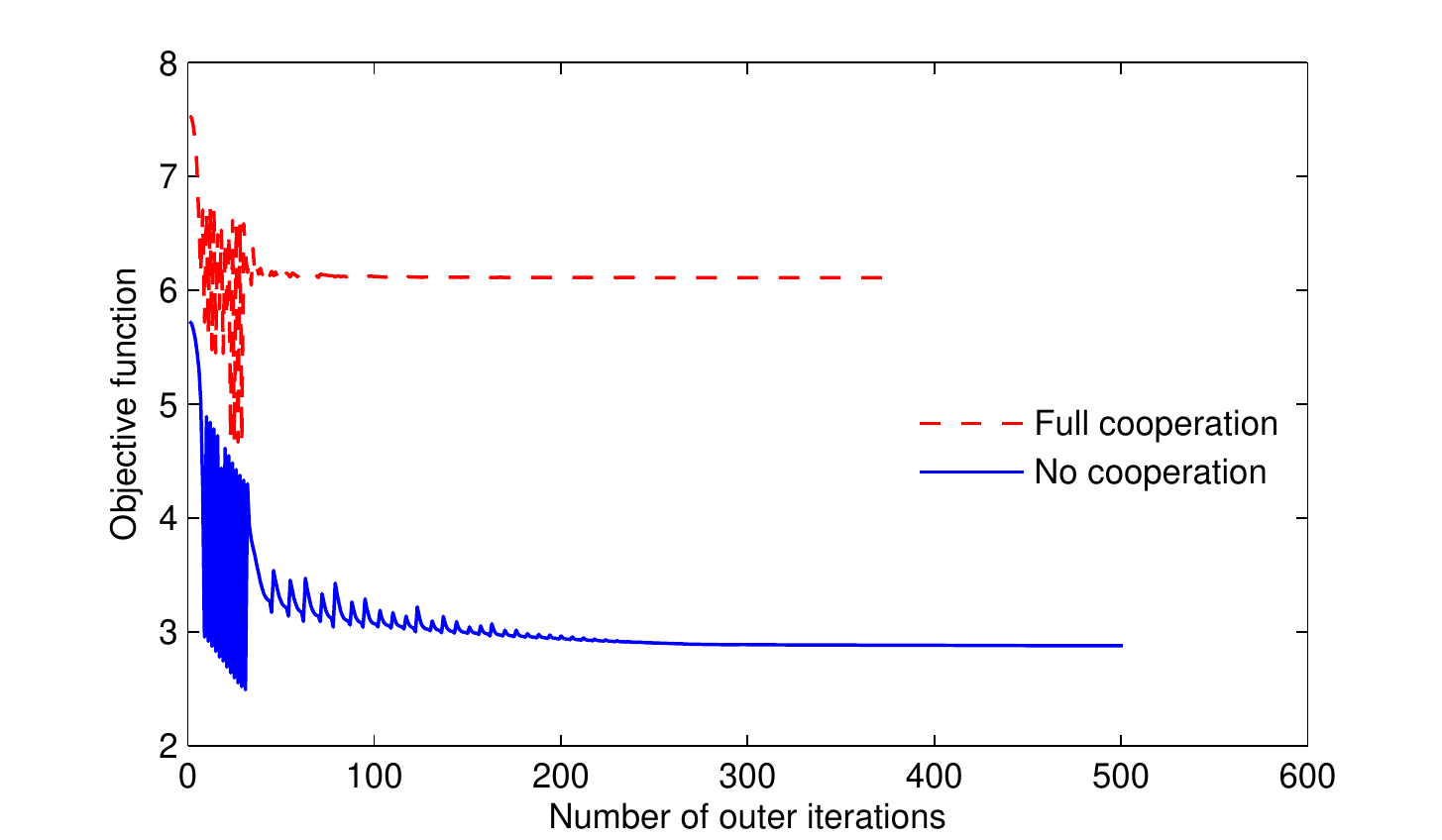}
\label{subfig:2cell-hfs-conv}
}
\subfigure[Individual group rates]{
\includegraphics[width=4.5in]{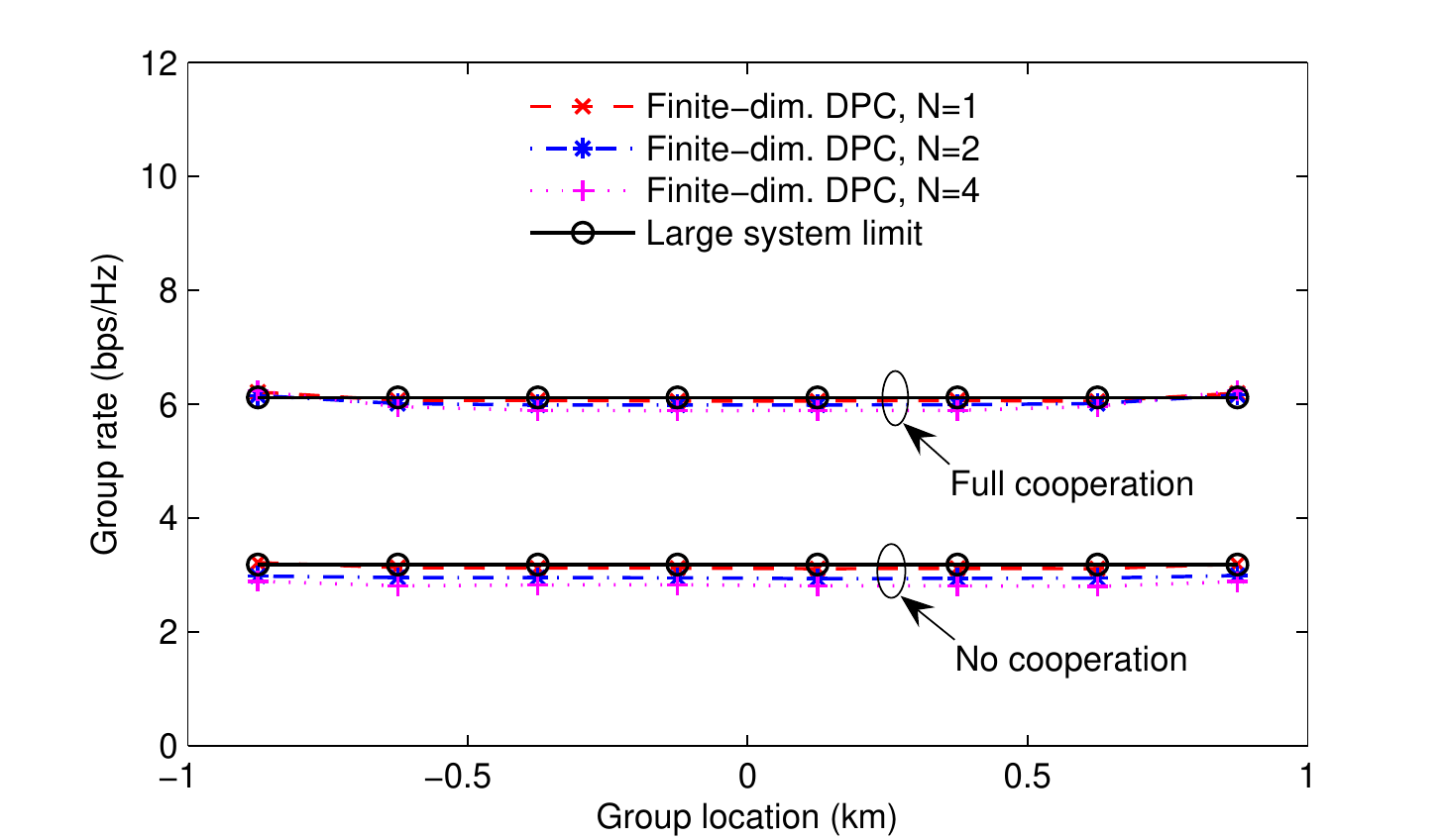}
\label{subfig:2cell-hfs-rate}
}
\caption{Hard fairness scheduling with $\gamma=4$ and $K=8$ in the 2-cell model.}
\label{fig:2cell-hfs}
\end{figure}

Fig. \ref{fig:2cell-hfs} shows the convergence behavior of the utility function and individual group
under HFS. In the HFS case, all the users achieve the same individual rate which is slightly higher than
the smallest rate of the PFS case. Also, the agreement with of the individual user rates with the finite
dimensional simulation is remarkable.

\begin{figure}
\centering
\subfigure[3-sectored cell configuration]{
\includegraphics[width=3in]{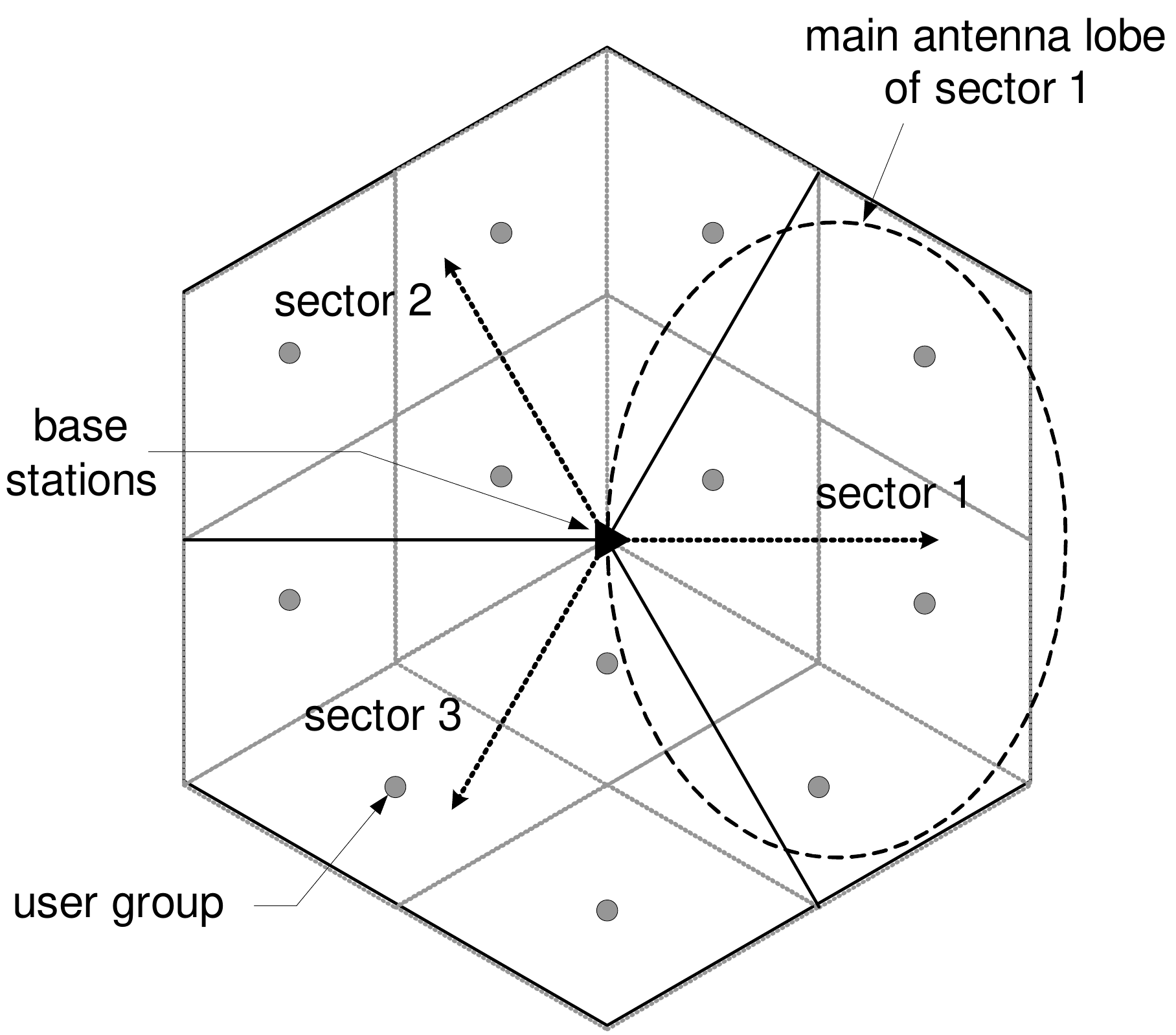}
\label{subfig:7cell-3sector}
}
\subfigure[Wrap-around torus topology]{
\includegraphics[width=2.7in]{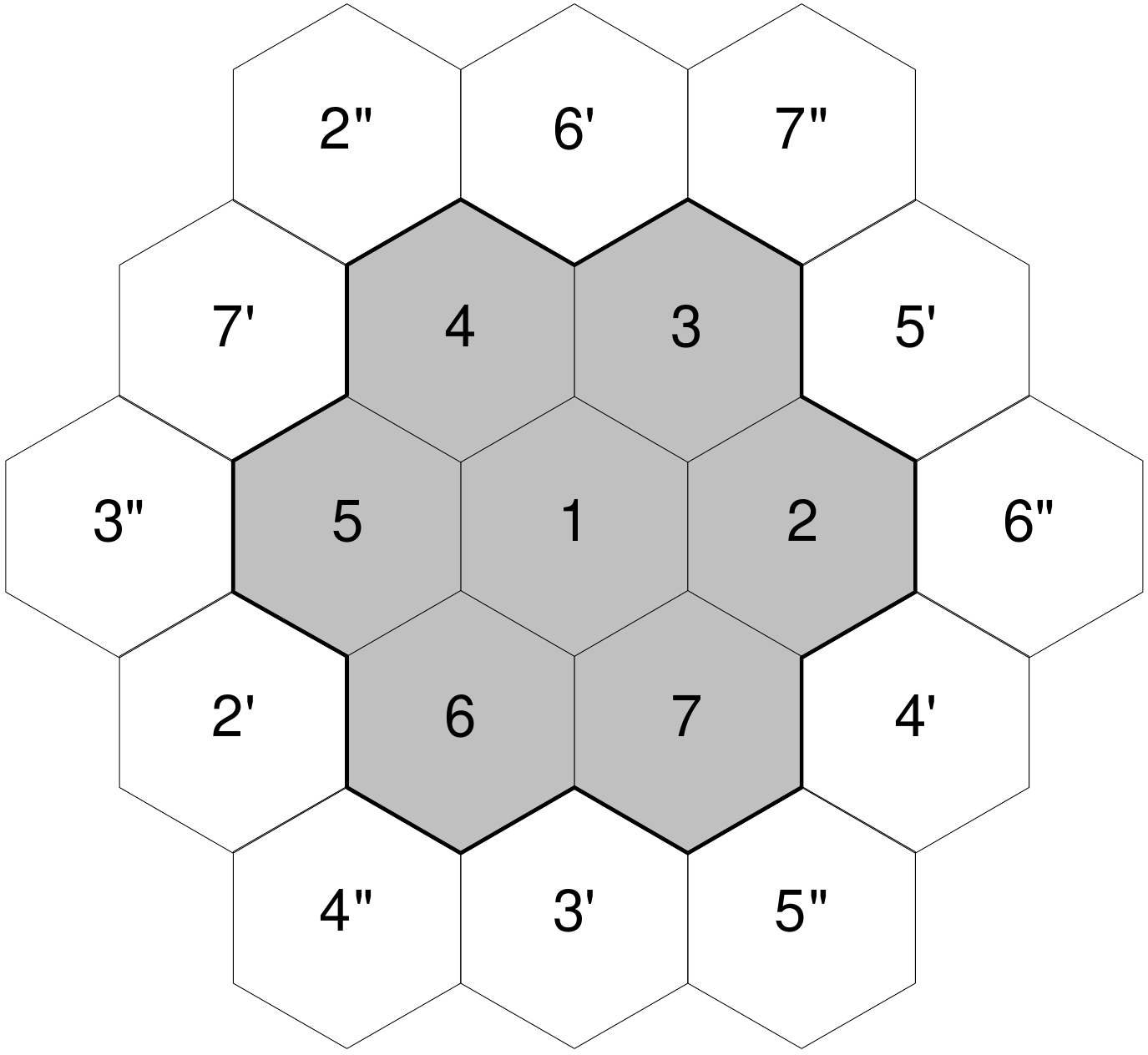}
\label{subfig:7cell-topol}
}
\caption{Two-dimensional three-sectored 7-cell model.}
\label{fig:7cell-model}
\end{figure}

\begin{figure}
\centering
\subfigure[No cooperation]{
\includegraphics[width=4.5in]{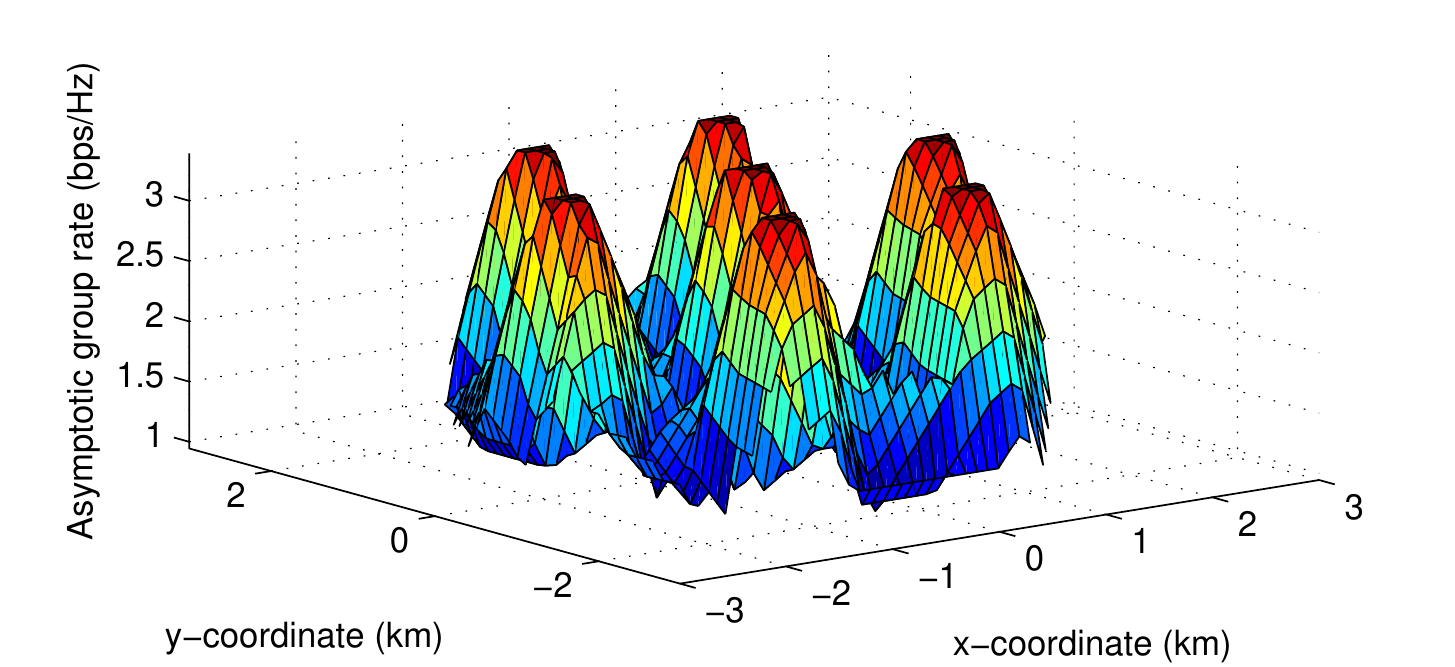}
\label{subfig:7cell-nocoop}
}
\subfigure[Cooperation among co-located sectors]{
\includegraphics[width=4.5in]{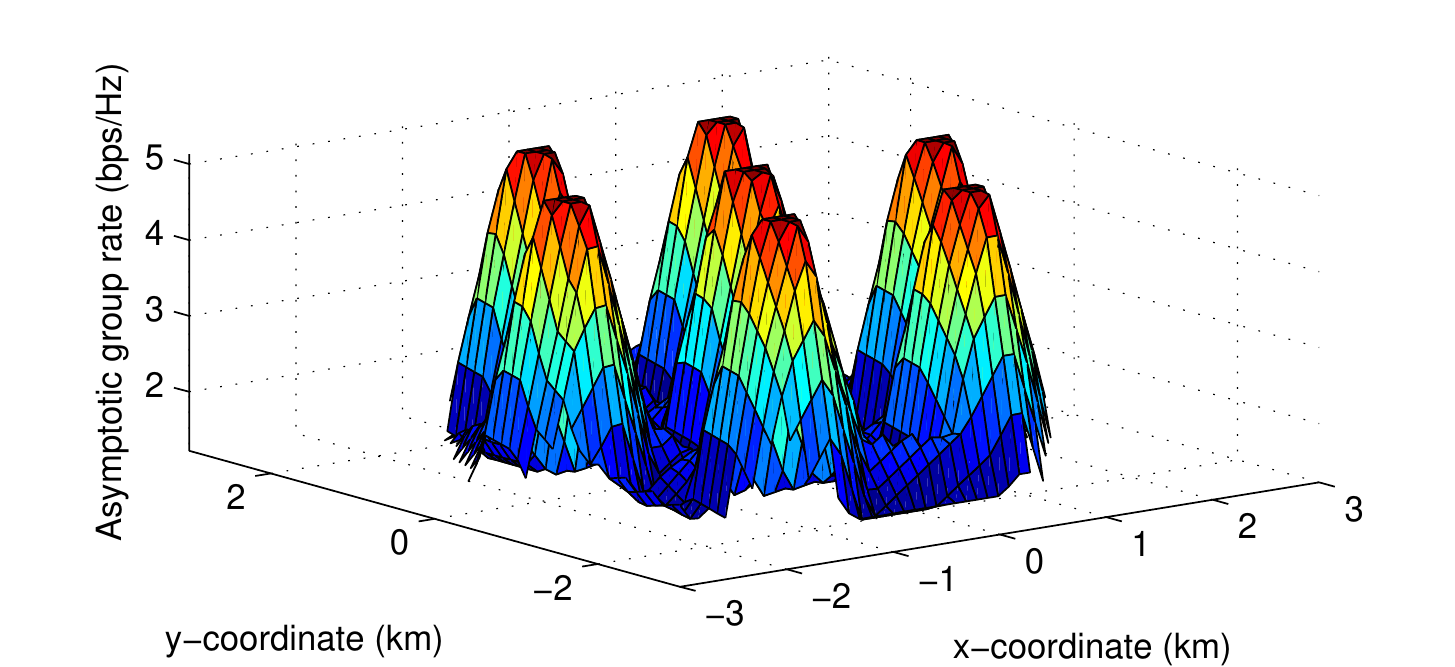}
\label{subfig:7cell-sctrcoop}
}
\subfigure[Full cooperation over 7 cells]{
\includegraphics[width=4.5in]{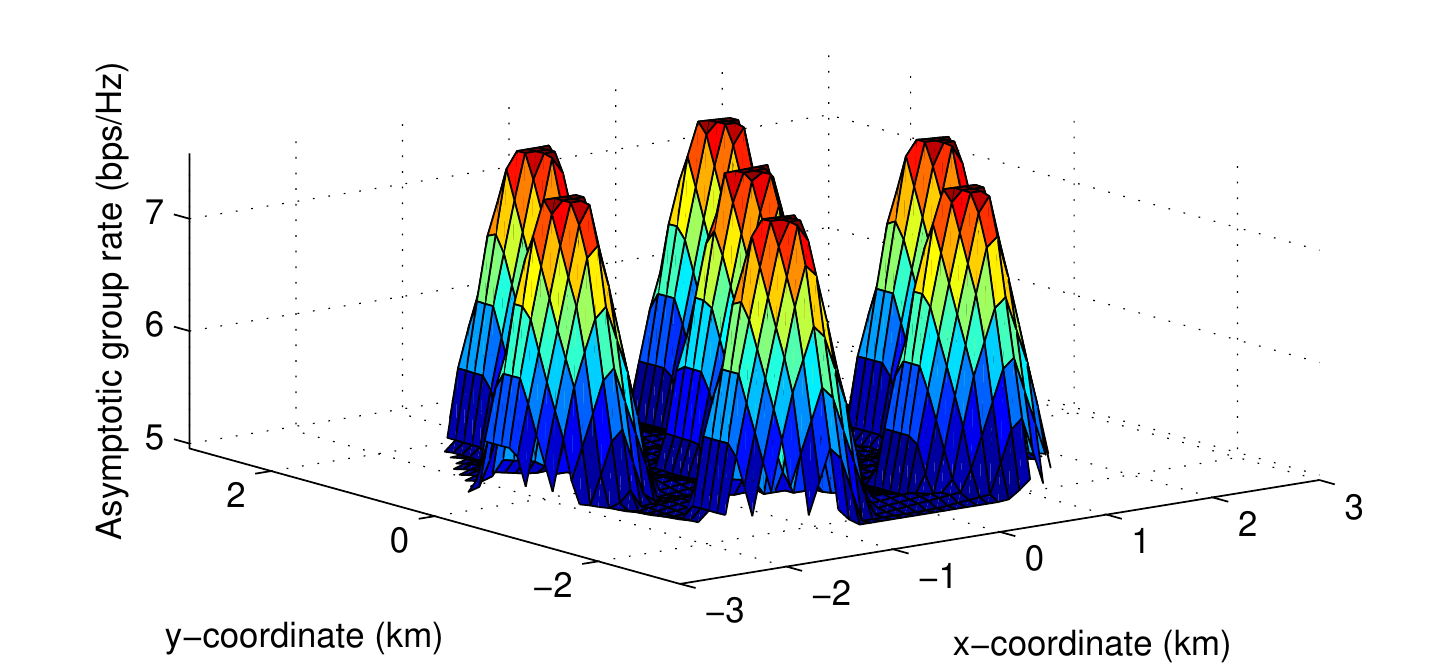}
\label{subfig:7cell-fullcoop}
}
\caption{Ergodic user rate distribution in the 7-cell model.}
\label{fig:7cell-pfs-rate}
\end{figure}

Using the proposed asymptotic analysis, validated in the simple 2-cell model, we can obtain ergodic
rate distributions for much larger systems, for which a full-scale simulation would be very demanding.
We considered a two-dimensional cell layout where 7 hexagonal cells form a network and each cell consists
of three 120-degree sectors. As depicted in Fig. \ref{subfig:7cell-3sector}, three BSs are co-located
at the center of each cell such that each BS handles one sector in no cooperation case. Each sector is
split into the 4 diamond-shaped equal-area grids and one user group is placed at the center of each grid.
Therefore there are total $M=21$ BSs and $K=84$ user groups in the network. In addition, we assume a
wrap-around torus topology as shown in Fig. \ref{subfig:7cell-topol}, such that each cell is virtually
surrounded by the other 6 cells and all the cells have the symmetric ICI distribution. The antenna
orientation and pattern follows \cite{IEEE80216m-EMD09} and the non-ideal spatial antenna gain pattern
(overlapping between sectors in the same cell) generates ICI even between sectors in the same cell
with no cooperation. This model is relevant for a macro-cell network where both the ICI and the effective
inter-cell cooperation are due to neighboring cells. Fig. \ref{fig:7cell-pfs-rate} shows the user rate
distribution under three levels of cooperation,
(a) no cooperation ($L=21$ single-sector clusters),
(b) cooperation among the co-located 3 sector BSs ($L=7$ clusters formed by three sectors of each cell),
and (c) full cooperation over 7-cell network ($L=1$).
From the asymptotic rate results, it is shown that in case (b), the cooperation gain over the case (a)
is primarily obtained for the users around cell centers, while the cooperation gain is attained over
the whole cellular coverage area in case (c).

\section{Conclusions} \label{sec:conclusion}

We considered the downlink of a multi-cell MU-MIMO cellular system where the pathloss and inter-cell
interference make the users' channel statistics unequal. In this case, it is important to
evaluate the system performance subject to some form of fairness. Downlink scheduling that make the
system operate at a desired point of the long-term average achievable rate region is an important
issue, widely studied and widely applied in practice \cite{Viswanath-Tse-Laroia-TIT02,
Bender-Viterbi-etal-CommMag00, Parkvall-Englund-Lundevall-Torsner-CommMag06}.
This is classically formulated as the maximization of a concave network utility function over the
achievable ergodic rate region of the system. We also considered an inter-cell cooperation scheme
for which groups of cells operates jointly, as a distributed multi-antenna transmitter, and have
perfect channel state information for all users in their cluster and only statistical information
on the inter-cluster interference. Under the constraint that inter-cluster interference is treated
as noise, this model is quite general.

We focused on the large-system limit where the number of base station antennas and the number of
users at each location go to infinity with a fixed ratio. In this regime, we presented a
semi-analytic method for the computation of the optimal fairness rate point, based on a combination
of large random matrix results and Lagrangian optimization. The proposed method is particularly
simple and efficient in the case where the system has certain symmetries. Otherwise, we can obtain
a simple upper bound by relaxing the per-base station power constraint to the per-cluster sum-power
constraint. Numerical results showed that the rates predicted by the large-system analysis are
indeed remarkably close to the rates by Monte Carlo simulation of a corresponding finite-dimensional
system. Overall, the results of this paper are useful in evaluating and optimizing the multi-cell
MU-MIMO systems, especially when the system dimension and network size are large.

\bibliography{tit-mimo-asympt-fairness-v1.3.bbl}

\end{document}

%% file: macros.tex
\long\def\comment#1{}

\newfont{\bbb}{msbm10 scaled 700}

\newfont{\bb}{msbm10 scaled 1100}
\newcommand{\CC}{\mbox{\bb C}}

\newcommand{\EE}{\mbox{\bb E}}

\newcommand{\av}{{\bf a}}

\newcommand{\nv}{{\bf n}}

\newcommand{\rv}{{\bf r}}
\newcommand{\sv}{{\bf s}}

\newcommand{\vv}{{\bf v}}
\newcommand{\xv}{{\bf x}}
\newcommand{\yv}{{\bf y}}
\newcommand{\zv}{{\bf z}}
\newcommand{\zerov}{{\bf 0}}

\newcommand{\Am}{{\bf A}}

\newcommand{\Gm}{{\bf G}}
\newcommand{\Hm}{{\bf H}}
\newcommand{\Id}{{\bf I}}

\newcommand{\Qm}{{\bf Q}}
\newcommand{\Rm}{{\bf R}}

\newcommand{\Tm}{{\bf T}}
\newcommand{\Um}{{\bf U}}
\newcommand{\Wm}{{\bf W}}

\newcommand{\Xm}{{\bf X}}

\newcommand{\Cc}{{\cal C}}

\newcommand{\Fc}{{\cal F}}
\newcommand{\Gc}{{\cal G}}

\newcommand{\Kc}{{\cal K}}
\newcommand{\Lc}{{\cal L}}
\newcommand{\Mc}{{\cal M}}
\newcommand{\Nc}{{\cal N}}

\newcommand{\Qc}{{\cal Q}}

\newcommand{\betav}{\hbox{\boldmath$\beta$}}

\newcommand{\lambdav}{\hbox{\boldmath$\lambda$}}
\newcommand{\epsilonv}{\hbox{\boldmath$\epsilon$}}

\newcommand{\Sigmam}{\hbox{\boldmath$\Sigma$}}

\newcommand{\Pim}{\hbox{\boldmath$\Pi$}}

\newcommand{\Thetam}{\hbox{\boldmath$\Theta$}}

\newcommand{\diag}{{\hbox{diag}}}

\newcommand{\trace}{{\hbox{tr}}}

\newcommand{\eqdef}{\stackrel{\Delta}{=}}

\newcommand{\herm}{{\sf H}}

\newcommand{\transp}{{\sf T}}

\newcommand{\Psf}{{\sf P}}
